\documentclass[%
 reprint,
 superscriptaddress,
 amsmath,amssymb,amsfonts,
 aps,
 pra,
 floatfix,
 longbibliography,nobibnotes
]{revtex4-2}

\usepackage{graphicx}
\usepackage{xcolor}
\usepackage{mathtools}

\usepackage[colorlinks=true]{hyperref}
\hypersetup{
  colorlinks=true,
  linkcolor=blue,
  urlcolor=cyan,
}

\usepackage{microtype}

\begin{document}

\title{Scaling and optimal preparation of spin-squeezed states with engineered gap protection}

\author{Arman Duha}
\affiliation{Department of Physics, Oklahoma State University, Stillwater, Oklahoma 74078, USA}

\author{Lucas Cartwright}
\affiliation{Department of Physics, Oklahoma State University, Stillwater, Oklahoma 74078, USA}

\author{Thomas Bilitewski}
\email{thomas.bilitewski@okstate.edu}
\affiliation{Department of Physics, Oklahoma State University, Stillwater, Oklahoma 74078, USA}

\date{\today}

\begin{abstract}
Spatially nonuniform interactions generate noncollective excitations that compete with and limit the achievable collective spin squeezing.
Heisenberg interactions suppress these excitations while leaving the collective squeezing dynamics intact.
We study this competition in single-layer two-axis-twisting and bilayer two-mode-squeezing models with local or power-law couplings in one and two dimensions.
We find that the achievable squeezing is controlled by the fraction $f_m$ of unstable finite-momentum modes.
Increasing the relative strength of the gap-protecting Heisenberg interactions reduces this fraction and improves the squeezing.
Scaling the protection strength with system size therefore yields a family of models with a tunable squeezing exponent in the partially collective regime between the standard quantum limit and the Heisenberg limit.
Finally, when considering the trade-off between stronger gap protection and slower entanglement dynamics at fixed total preparation time, intermediate protection can yield better squeezing, even when stronger protection would eventually reach a deeper minimum.
In this regime, the best squeezing at fixed preparation time improves algebraically rather than exponentially with available time.
The control of spatial fluctuations thus determines both how squeezing scales and how quickly strongly entangled states can be prepared.
\end{abstract}

\maketitle

\section{Introduction}
\label{sec:introduction}

Power-law interacting spin systems realized in Rydberg atoms~\cite{Browaeys2020}, polar molecules~\cite{Bohn2017}, magnetic atoms~\cite{Chomaz2023}, and trapped ions~\cite{Monroe2021} enable controlled studies of entanglement generation in many-body dynamics.
Optical tweezers and lattices combined with single-particle control allow the preparation of ordered arrays, bilayers, and multilayers~\cite{Barredo2018,Tobias2022,Du2024,PhysRevLett.130.180601,Hawaldar2024,PhysRevResearch.7.L022045}, while Floquet pulse sequences engineer the spin structure of the native interactions~\cite{Lukin_2020_Robust,Geier2021,Scholl2022,PRXQuantum.4.010334,PhysRevA.108.053318,Tian_2025}.
In particular, Floquet engineering in polar molecules has realized two-axis-twisting mean-field dynamics from polarized initial states~\cite{Miller2024}, motivating the study of the entangled states accessible under such interactions.

Spin squeezing is a widely used form of entanglement for quantum metrology~\cite{RevModPhys.89.035002,RevMod_Metrology_2018,Ma2011,montenegro2024reviewquantummetrologysensing}.
By reducing the fluctuations of one collective-spin quadrature while retaining a large mean spin, squeezed states can improve sensitivity beyond the standard quantum limit~\cite{Wineland1992,wineland1994}, with demonstrated gains in atomic clocks~\cite{pedrozo2020entanglement,Robinson2024} and magnetometers~\cite{PhysRevLett.109.253605,PhysRevLett.113.103004}.
While fully exploiting these squeezed states requires single-particle detection, time-reversal readout protocols alleviate this requirement~\cite{PhysRevLett.116.053601,Colombo2022}.
In collective spin models, where every particle couples equally to all others, one-axis twisting (OAT) produces squeezing that scales as $N^{-2/3}$, while two-axis twisting (TAT) reaches the Heisenberg scaling $N^{-1}$~\cite{Kitagawa1993,Kajtoch2015}, motivating proposals for engineering TAT dynamics~\cite{Liu2011,Borregaard2017,Groszkowski2020}.
Because TAT squeezes exponentially in time, it reaches its optimum in a time that grows only logarithmically with $N$, making it well suited for preparing strongly entangled states rapidly.
Whether this collective behavior survives for spatially decaying interactions is a central question for extended spin systems.

Finite-range interactions couple the collective spin to spatially nonuniform excitations, which generally limit the achievable squeezing.
Nevertheless, scalable squeezing does not require uniform all-to-all interactions~\cite{Rey2008,PhysRevLett.125.223401,fossfeig2016entanglementspinsqueezinginfiniterangeinteractions}.
Spin-exchange Heisenberg interactions can protect collective behavior through an energy gap~\cite{Rey2008,PhysRevLett.125.223401} even in short-range or disordered interacting systems.
Squeezing with finite-range interactions has now been demonstrated in trapped ions, Rydberg arrays, polar molecules, gases with contact interactions, and solid-state dipolar spin ensembles~\cite{Bornet2023a,Franke2023a,Eckner2023,Hines2023,wu2025spinsqueezingensemblenitrogenvacancy,jqvz-kpwg,holland2026creatingprobingspinsqueezedstates}.
This persistence of collective dynamics has been explained in terms of the low-energy tower of states, spontaneous continuous symmetry breaking, and the separation of collective rotation from spin waves~\cite{Roscilde2022,Comparin2022PRL,Roscilde2023}.
From a different point of view, scalable squeezing has been connected to the presence of finite-temperature easy-plane order~\cite{Block_2024}, to critical slowing down in systems with quasi-long-range order~\cite{Roscilde2024}, and has been predicted to be robust to the presence of disorder~\cite{kaplanlipkin2025theoryscalablespinsqueezing,begg2026scalable,solfanelli2026robustspinsqueezingquantumnetworks}.
In those OAT, XY, and XXZ models, spin waves remain dynamically stable around the polarized state.
In contrast, for the pair-creation dynamics considered here, an extensive set of finite-momentum modes is exponentially unstable unless suppressed.
For TAT, combining the squeezing interaction with Heisenberg exchange has recently been proposed as a route toward achieving the Heisenberg limit with power-law or short-range interacting spins~\cite{koyluoglu2025squeezingheisenberglimitlocally}.
More generally, gap protection can be extended to generic squeezing mechanisms and multilevel qudit systems~\cite{Bahamnia2026}.

Bilayers offer a related mechanism in which exchange between oppositely polarized layers creates correlated pairs of spin excitations~\cite{Bilitewski2023,Bilitewski2023a,Dominguez2024}.
This process generates bosonic two-mode squeezing (TMS)~\cite{Caves1985,Schumaker1985}, analogous to spin-mixing and cavity-mediated pair creation in atomic ensembles~\cite{Gross2011,Lange2018,Fadel2018,Kunkel2018,Sundar2023}, reducing fluctuations of joint spin quadratures of the two layers.
By engineering Heisenberg interactions within each layer and spin exchange between them, Heisenberg-limited squeezing can be obtained with power-law interactions~\cite{PhysRevA.109.L061304}.
The intralayer exchange detunes nonuniform excitations while leaving the uniform pair-production instability intact.
The competition between this protection and the nonuniform interlayer coupling results in a transition from a fully collective to a partially collective regime with multiple unstable finite-momentum modes~\cite{7g55-lpff}.
This transition and its critical scaling were found to be universal across different lattice geometries and changes of the relative coupling strengths~\cite{Duha2026PRA}.

These works identified the transition but did not determine what controls squeezing within the partially collective regime or how its scaling can be tuned.
Suppressing every noncollective instability requires a protection strength that typically grows with system size, whereas a less restrictive regime permits multiple unstable modes.
This raises the question of whether squeezing can still improve with system size in the presence of multiple unstable modes, and how its scaling depends on them.

\begin{figure}[tbp]
 \centering
 \includegraphics[width=\columnwidth]{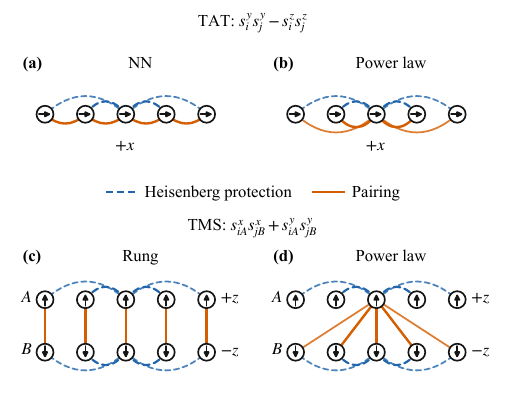}
\caption{Spin models and interaction geometries. (a), (b) Single-layer TAT with nearest-neighbor (NN) and power-law pairing. (c), (d) Bilayer TMS with rung and power-law interlayer pairing.
Blue dashed bonds denote power-law Heisenberg protection within each layer, orange solid bonds denote the pair-generating interactions, whose spin structure is indicated above each row.
Arrows show the initial polarizations: $+x$ for TAT, and $+z$ ($-z$) in layer $A$ ($B$) for TMS.
Only representative bonds are shown, with line weights indicating decay qualitatively.
The chain drawings illustrate models studied on periodic chains and square lattices.}
 \label{fig:interaction_schematic}
\end{figure}

Here we investigate this question in single-layer TAT and bilayer TMS models with tunable Heisenberg protection and either local or power-law squeezing interactions, illustrated in Fig.~\ref{fig:interaction_schematic}.
Combining Bogoliubov theory with discrete truncated Wigner approximation (dTWA) simulations, we find that the squeezing minimum follows a power law in the fraction $f_m$ of unstable finite-momentum modes, rather than being set by their number.
Squeezing therefore improves with system size whenever $f_m$ decreases, even as the number of unstable modes grows; suppressing every nonuniform instability is unnecessary, and scaling the protection strength with system size yields a continuously tunable squeezing exponent.

When the gap protection is engineered from the same interaction resource as the squeezing, stronger protection also slows the squeezing dynamics.
For one-dimensional systems, we find that the optimal protection depends on the available preparation time and generally lies below the fully protected limit; the best squeezing then improves only algebraically with time rather than exponentially.

In Sec.~\ref{sec:model} we introduce the spin models, the Bogoliubov analysis, and the numerical methods.
Sections~\ref{sec:mode_fraction} and \ref{sec:composition} demonstrate the relation of achievable squeezing to the unstable-mode fraction and study its control through size-dependent protection.
Section~\ref{sec:time} examines the resulting optimal preparation protocols, and Sec.~\ref{sec:outlook} concludes.

\section{Model}
\label{sec:model}

\subsection{Spin models}
We consider spin-$1/2$ systems on periodic lattices in single-layer and bilayer geometries whose all-to-all limits are TAT and TMS (Fig.~\ref{fig:interaction_schematic}).
We study chains and ladders in $d=1$, and square layers and bilayers in $d=2$, with linear extent $L$ and unit lattice spacing, containing $N=L^d$ spins for TAT and $N=2L^d$ total spins ($L^d$ per layer) for TMS.
We set $\hbar=1$ and define the site spin operators $\hat s_i^\mu=\hat\sigma_i^\mu/2$.

The two models we consider are described by
\begin{align}
 \hat H_{\mathrm{TAT}}=-2 \sum_{i<j}\bigl[ &V_{ij}(\hat s_i^y\hat s_j^y-\hat s_i^z\hat s_j^z)
 + \chi W_{ij} \,\hat{\boldsymbol s}_i\cdot\hat{\boldsymbol s}_j\bigr],
 \label{eq:tat_hamiltonian}
\end{align}
and
\begin{align}
 \hat H_{\mathrm{TMS}}={}& \sum_{i,j} V_{ij}(\hat s_{iA}^x\hat s_{jB}^x+\hat s_{iA}^y\hat s_{jB}^y)\nonumber\\
 &+ \chi \sum_{\eta=A,B}\sum_{i<j} W_{ij} \,
 \hat{\boldsymbol s}_{i\eta}\cdot\hat{\boldsymbol s}_{j\eta},
 \label{eq:tms_hamiltonian}
\end{align}
where $i,j$ enumerate sites in a chain or layer, and $\eta\in\{A,B\}$ distinguishes the legs of the ladder or layers of the bilayer.

For both Hamiltonians, the first term generates squeezing through pair creation, while the second is a Heisenberg interaction that protects the collective dynamics with an energy gap.
The prefactor $\chi$ tunes the protection strength relative to the pair-creation term~\cite{PhysRevA.109.L061304,7g55-lpff,Duha2026PRA}.
The prefactors in Eqs.~\eqref{eq:tat_hamiltonian} and \eqref{eq:tms_hamiltonian} fix the time units, giving the collective squeezing rates $\Gamma_0=2\Omega_0$ (TAT) and $\Omega_0$ (TMS) derived in Sec.~\ref{sec:bogoliubov}.

We assume power-law interactions for the Heisenberg term,
\begin{equation}
  W_{ij}=r_{ij}^{-\alpha},\quad i\ne j,
 \label{eq:protection_kernel}
\end{equation}
with $W_{ii}=0$ (and $V_{ii}=0$ for TAT), where $r_{ij}$ is the periodic minimum-image distance within a layer and $\alpha$ sets the power-law decay exponent.
We compare two choices for the pair-creation interaction $V_{ij}$ in each geometry: for TAT, nearest-neighbor (NN) pairing and $V_{ij}\propto r_{ij}^{-\alpha}$ [Figs.~\ref{fig:interaction_schematic}(a) and \ref{fig:interaction_schematic}(b)]; for TMS, rung pairing ($V_{ij}\propto\delta_{ij}$) and $V_{ij}\propto(r_{ij}^2+a_Z^2)^{-\alpha/2}$ at fixed layer separation $a_Z=1$ [Figs.~\ref{fig:interaction_schematic}(c) and \ref{fig:interaction_schematic}(d)].
We label the four resulting models TAT--NN, TAT--power law, TMS--rung, and TMS--power law.
We normalize the interactions such that $\Omega_0=\sum_j V_{ij}=1$, so $\chi$ controls protection relative to the collective pair-creation scale.
A common rescaling of both interactions changes only the time scale.
The conversion from unnormalized couplings is given in Appendix~\ref{app:v:budgets}.

For the single-layer TAT model, we prepare a coherent spin state polarized along $+x$, with $\langle\hat{\boldsymbol S}\rangle=(N/2)\hat{\boldsymbol x}$ and $\hat{\boldsymbol S}=\sum_i\hat{\boldsymbol s}_i$.
For TMS, we initialize oppositely polarized layers, $\langle\hat{\boldsymbol S}_A\rangle=-\langle\hat{\boldsymbol S}_B\rangle=(L^d/2)\hat{\boldsymbol z}$, where $\hat{\boldsymbol S}_\eta=\sum_i\hat{\boldsymbol s}_{i\eta}$.
Both initial states are eigenstates of the Heisenberg terms, so their evolution is driven solely by the pair-creation terms.
The TAT anisotropy creates two flips relative to $+x$, while interlayer exchange creates one excitation relative to each layer's polarization, conserving the total $z$ magnetization.
At the same time, Heisenberg interactions detune nonuniform excitations without affecting the collective excitations that generate squeezing.

\subsection{Bogoliubov analysis and collectivity}
\label{sec:bogoliubov}

In momentum space, the competition between pair creation and protection can be understood through a Bogoliubov analysis.
We begin with a Holstein--Primakoff transformation~\cite{Holstein1940}, expressing spin fluctuations about the initial state in terms of bosonic operators.
For TAT, the quantization axis is $+x$.
With $\hat s_{i,x}^{\pm}=\hat s_i^y\pm i\hat s_i^z$, to leading order
\begin{equation}
 \hat s_i^x=\tfrac12-\hat a_i^\dagger\hat a_i,
 \qquad \hat s_{i,x}^+\simeq\hat a_i,\qquad \hat s_{i,x}^-\simeq\hat a_i^{\dagger}.
 \label{eq:hp_tat}
\end{equation}
For TMS, using $\hat s_{i\eta}^{\pm}=\hat s_{i\eta}^x\pm i\hat s_{i\eta}^y$, expanding around the oppositely polarized layers gives
\begin{align}
 \hat s_{iA}^z&=\tfrac12-\hat a_i^\dagger\hat a_i,
 &\hat s_{iA}^+&\simeq\hat a_i,\nonumber\\
 \hat s_{iB}^z&=-\tfrac12+\hat b_i^\dagger\hat b_i,
 &\hat s_{iB}^+&\simeq\hat b_i^\dagger.
 \label{eq:hp_tms}
\end{align}
Thus, interlayer spin exchange maps to bosonic pair creation and annihilation.

Transforming to momentum space and keeping terms to quadratic order yields, up to additive constants,
\begin{align}
 \hat H_{\mathrm{TAT}}^{(2)}&=\sum_{\boldsymbol k}\Bigl[\chi\Delta_{\boldsymbol k}\hat a_{\boldsymbol k}^\dagger\hat a_{\boldsymbol k}
 -\tfrac12\Omega_{\boldsymbol k}\bigl(\hat a_{\boldsymbol k}\hat a_{-\boldsymbol k}+\mathrm{H.c.}\bigr)\Bigr],
 \nonumber\\
 \hat H_{\mathrm{TMS}}^{(2)}&=\sum_{\boldsymbol k}\Bigl[-\tfrac12\chi\Delta_{\boldsymbol k}\bigl(\hat a_{\boldsymbol k}^\dagger\hat a_{\boldsymbol k}+\hat b_{\boldsymbol k}^\dagger\hat b_{\boldsymbol k}\bigr)
 \nonumber\\
 &\qquad+\tfrac12\Omega_{\boldsymbol k}\bigl(\hat a_{\boldsymbol k}\hat b_{-\boldsymbol k}+\mathrm{H.c.}\bigr)\Bigr],
 \label{eq:quadratic_hamiltonians}
\end{align}
in terms of the Fourier transforms
\begin{align}
 W_{\boldsymbol k}&=\sum_{\boldsymbol r\ne0}W_{\boldsymbol r}e^{i\boldsymbol k\cdot\boldsymbol r},
 \qquad \Delta_{\boldsymbol k}=W_{\boldsymbol 0}-W_{\boldsymbol k},
 \nonumber\\
 \Omega_{\boldsymbol k}&=\sum_{\boldsymbol r}V_{\boldsymbol r}e^{i\boldsymbol k\cdot\boldsymbol r},
 \label{eq:spectra}
\end{align}
where $\chi\Delta_{\boldsymbol k}$ sets the energy gap for pair creation, $\Omega_{\boldsymbol k}$ is the pairing strength, and the pairing sum omits $\boldsymbol r=0$ for TAT while including $\boldsymbol r=0$ (rung coupling) for TMS.

The Bogoliubov eigenenergies satisfy
\begin{align}
 \omega_{\boldsymbol k,\mathrm{TAT}}^2
 &=\chi^2\Delta_{\boldsymbol k}^2-|\Omega_{\boldsymbol k}|^2,\nonumber\\
 \omega_{\boldsymbol k,\mathrm{TMS}}^2
 &=\tfrac14\bigl(\chi^2\Delta_{\boldsymbol k}^2-|\Omega_{\boldsymbol k}|^2\bigr).
 \label{eq:bogoliubov}
\end{align}
For modes satisfying
\begin{equation}
 |\Omega_{\boldsymbol k}|>\chi\Delta_{\boldsymbol k},
 \label{eq:instability}
\end{equation}
the energies become imaginary, leading to exponential growth of fluctuations and mode occupations.
The collective $\boldsymbol k=\boldsymbol 0$ mode is always unstable because $\Delta_{\boldsymbol 0}=0$ and $\Omega_0\ne0$.
Its squeezed variance initially decays as $\exp(-\Gamma_0 t)$, where $t$ is the time in the units of Eqs.~\eqref{eq:tat_hamiltonian} and \eqref{eq:tms_hamiltonian} with $\Omega_0=1$, and $\Gamma_0=2\Omega_0$ for TAT while $\Gamma_0=\Omega_0$ for TMS.

\begin{figure}[tbp]
 \centering
 \includegraphics[width=\columnwidth]{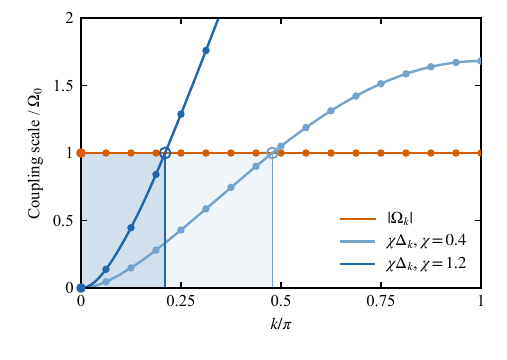}
\caption{Competition between pairing and Heisenberg protection in the Bogoliubov analysis.
The pairing strength $|\Omega_k|=1$ and energy gap $\chi\Delta_k$ are shown for rung TMS on a periodic chain with $L=32$ and $\alpha=3$.
Light and dark blue shading mark the unstable momentum intervals for $\chi=0.4$ and $1.2$, respectively.
Dots mark the allowed momenta $k=2\pi n/L$; open markers at the crossings indicate the instability boundaries $k_c$.
Stronger protection narrows the unstable region while preserving the collective instability at $k=0$.}
 \label{fig:bogoliubov_protection}
\end{figure}

The protection strength $\chi$ controls which modes are stable and unstable.
Figure~\ref{fig:bogoliubov_protection} shows the crossing of the pairing and protection interaction terms for rung pairing, where $|\Omega_k|=\Omega_0$ is independent of momentum.
Increasing $\chi$ moves the crossing to smaller $k$, suppressing more unstable modes.

On a finite lattice, the allowed momenta have components $k_\mu=2\pi n_\mu/L$.
We count each unstable nonzero pair $\{\boldsymbol k,-\boldsymbol k\}$ once and denote the count by $m$.
The corresponding unstable-mode fraction is
\begin{equation}
 f_m=\frac{m}{L^d},
 \label{eq:mode_fraction}
\end{equation}
excluding the collective mode (so $f_m$ counts unstable pairs per site and is bounded above by $\approx 1/2$).
We use $f_m$ to characterize the extent of the unstable region and test its relation to the achievable squeezing in Sec.~\ref{sec:mode_fraction}.

The quadratic theory predicts a collective-only instability~\cite{koyluoglu2025squeezingheisenberglimitlocally,7g55-lpff,Duha2026PRA} when
\begin{equation}
 \chi>\chi_c(L),\qquad
 \chi_c(L)=\max_{\boldsymbol k\ne\boldsymbol 0}\frac{|\Omega_{\boldsymbol k}|}{\Delta_{\boldsymbol k}}.
 \label{eq:critical_protection}
\end{equation}
Below this threshold, additional finite-momentum modes are unstable.
We refer to this regime as partially collective~\cite{7g55-lpff,Duha2026PRA}.

\subsection{Squeezing observables and numerical methods}
For TAT, we use the Wineland parameter~\cite{wineland1994},
\begin{equation}
 \xi^2_{\mathrm{TAT}}(t)= \frac{N \min_{\boldsymbol{n}_\perp} \operatorname{Var}(\boldsymbol{n}_\perp \cdot \hat{\boldsymbol{S}})}{|\langle\hat{\boldsymbol{S}}\rangle|^2},
 \label{eq:tat_observable}
\end{equation}
where $\boldsymbol{n}_\perp$ is a unit vector perpendicular to the mean spin $\langle\hat{\boldsymbol{S}}\rangle$.

For TMS, we minimize the variance of joint transverse spin components.
Writing $\hat{\boldsymbol Q} =(\hat S_A^x,\hat S_A^y,\hat S_B^x,\hat S_B^y)$, we form the covariance matrix
\begin{equation}
 \Sigma_{ab}
 =
 \frac12\left\langle
 \{\delta\hat Q_a,\delta\hat Q_b\}
 \right\rangle
\end{equation}
and define the Wineland-type parameter
\begin{equation}
 \begin{aligned}
 \xi^2_{\mathrm{TMS}}(t)
 &=
 \frac{4\lambda_{\min}[\Sigma(t)]}
 {L^d c_{\mathrm{av}}^2(t)},\\
 c_{\mathrm{av}}(t)
 &=
 \frac{
 |\langle\hat{\boldsymbol S}_A(t)\rangle|
 +
 |\langle\hat{\boldsymbol S}_B(t)\rangle|
 }{L^d}.
 \end{aligned}
 \label{eq:tms_observable}
\end{equation}
The smallest eigenvalue $\lambda_{\min}[\Sigma(t)]$ is the minimum variance among normalized linear combinations of $\hat{\boldsymbol Q}$, and $c_{\mathrm{av}}(t)$ measures the remaining contrast of the individual layers.
The parameter $\xi^2_{\mathrm{TMS}}$ quantifies the metrological phase sensitivity $(\Delta\theta)^2 = \xi^2_{\mathrm{TMS}}/N$ relative to the standard quantum limit $1/N$ for a transverse rotation matching the squeezed direction (e.g., generated by $\hat S_A^y - \hat S_B^x$ for the collective quadrature $(\hat S_A^x-\hat S_B^y)/\sqrt2$ in our conventions~\cite{PhysRevA.109.L061304,Duha2026PRA}).

We denote the minimum over the simulated preparation interval by $\xi^2_{\min}$ and characterize its size dependence by
\begin{equation}
 \xi^2_{\min}\propto N^{-u}.
 \label{eq:squeezing_size_scaling}
\end{equation}
Here $u=1$ corresponds to Heisenberg scaling of the squeezing parameter, $0<u<1$ gives a slower improvement with size, and $u=0$ gives size-independent squeezing. Squeezing in decibels is given by $10\log_{10}\xi^2$, with negative values indicating squeezing below the standard quantum limit.

To simulate the dynamics beyond the initial Bogoliubov regime, we use dTWA~\cite{Polkovnikov2010,Schachenmayer2015,Zhu_NewJournalofPhysics_21_2019}, a semiclassical phase-space approach.
Unless noted otherwise, individual dTWA runs use 2000 trajectories.
Statistical convergence of the minimum squeezing under dTWA is analyzed in Appendix~\ref{app:dtwa_sampling}.
While dTWA has been benchmarked against exact methods~\cite{Muleady_PRL_2023}, we provide model-specific comparisons against matrix-product-state (MPS) evolution in Appendix~\ref{app:mps} and against exact diagonalization (ED) in the collective limit in Appendices~\ref{app:v:tmscollective} and \ref{app:v:tatcollective}, and discuss the limitations of the semiclassical approximation.

\section{Squeezing controlled by the unstable-mode fraction}
\label{sec:mode_fraction}

\subsection{Protection strength and the unstable fraction}

Whereas previous studies used the competition between pair creation and Heisenberg protection to locate the boundary $\chi_c(L)$ between fully and partially collective dynamics~\cite{7g55-lpff,Duha2026PRA}, here we use $\chi$ to control the unstable-mode fraction $f_m$ within the partially collective regime.

For integrable interactions, $\alpha>d$, the long-wavelength dispersion obeys~\cite{RevModPhys.95.035002}
\begin{equation}
 \Delta_{\boldsymbol k}\propto |\boldsymbol k|^\nu,
 \qquad \nu=\min(\alpha-d,2),
 \label{eq:dispersion_scaling}
\end{equation}
for $\alpha\ne d+2$.
The nonanalytic exponent $\alpha-d$ reflects the long-range tail of the interactions.
In contrast, for sufficiently short-range interactions, when the second spatial moment of the interaction converges, the leading term is quadratic.
For $\Omega_{\boldsymbol k}\to\Omega_0\ne0$ at small $|\boldsymbol k|$ and large $\chi$, the instability condition~\eqref{eq:instability} gives
\begin{equation}
 k_c\propto\chi^{-1/\nu},\qquad
 f_m\propto k_c^d\propto\chi^{-d/\nu}.
 \label{eq:f_chi}
\end{equation}

At $\alpha=d+2$, the dispersion instead acquires a logarithm~\cite{RevModPhys.95.035002}.
For the one-dimensional $\alpha=3$ case, $\Delta_k=k^2[\ln(1/|k|)+3/2]+O(k^4)$~\cite{Duha2026PRA}.
Since $\ln(1/k_c)\approx2$--$4$ for the system sizes and protection strengths considered here, the constant $3/2$ is not negligible.
The threshold for suppressing finite-momentum instabilities scales as $\chi_c\propto L^2/\ln L$, rather than a pure $L^2$ power.
Away from this marginal case, $\chi_c\propto L^\nu$.

\begin{figure*}[tbp]
 \centering
 \includegraphics[width=\textwidth]{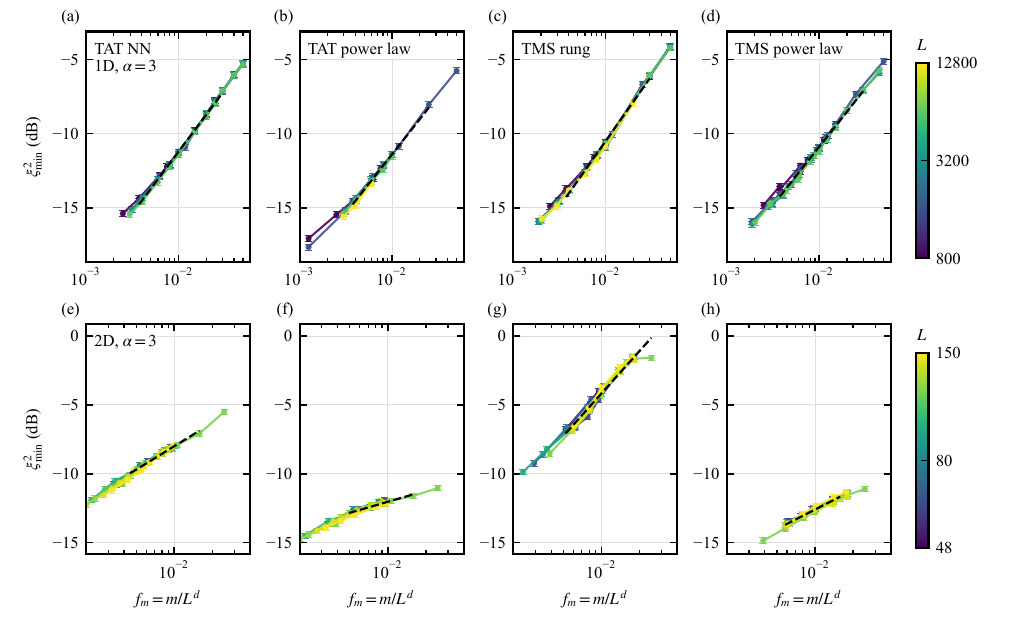}
\caption{Squeezing $\xi^2_{\min}$ controlled by the unstable-mode fraction at $\alpha=3$ in one dimension (upper row) and two dimensions (lower row), for all four models.
Colors distinguish linear sizes $L$, and thin lines connect fixed-size scans.
Dashed curves show $C f_m^q$ over the fitted interval, with $C$ and $q$ fitted separately for each panel (Table~\ref{tab:composition_coefficients} in Appendix~\ref{app:composition_calibration}).
Vertical bars show a common representative sampling uncertainty of $\pm0.2\,\mathrm{dB}$, based on the 2000-trajectory one-dimensional convergence benchmarks in Appendix~\ref{app:dtwa_sampling}.}
 \label{fig:mode_fraction}
\end{figure*}

\subsection{Squeezing organized by the mode fraction}

Figure~\ref{fig:mode_fraction} shows the minimum squeezing in chains and square lattices with both local and power-law pairing: across the explored range, data for different system sizes collapse approximately onto a single curve that deepens as $f_m$ decreases.
At fixed $f_m$, a larger system does not squeeze more strongly; scalable squeezing is only achievable if the fraction $f_m$ decreases with system size.
The corresponding one-dimensional scans at other interaction exponents are shown in Appendix~\ref{app:composition_calibration}.

We describe this dependence over the tested partially collective domain by
\begin{equation}
 \xi^2_{\min}(L,\chi)
 =C\,[f_m(L,\chi)]^q.
 \label{eq:empirical_composition}
\end{equation}
For each model, pairing interaction, dimension, and $\alpha$, the same $C$ and $q$ are used for all system sizes and protection strengths.
They are fitted to fixed-size scans over the $f_m$ interval spanned by accessible system sizes and gap protection strengths.
Appendix~\ref{app:composition_calibration} provides full details and lists all fitted coefficients in Table~\ref{tab:composition_coefficients}.
The coefficient $C$ fixes the absolute squeezing scale, while $q$ measures its dependence on the unstable-mode fraction.

Equation~\eqref{eq:empirical_composition} applies within the partially collective regime ($m\ge1$), up to finite-size corrections that are largest for the smallest systems; as $f_m\to0$ it must cross over to the collective result $\xi^2_{\min}\propto1/N$.
Section~\ref{sec:composition} shows that it nevertheless predicts the size scaling of squeezing quantitatively.

The dependence of the collapse on the form of the interactions and model shows that counting modes alone is not sufficient to explain the full dynamics.
Models with the same number of unstable modes can have different couplings and growth rates, and the squeezing minimum involves nonlinear evolution beyond the initial Bogoliubov approximation.
Each unstable mode grows at rate $\gamma_{\boldsymbol k}=\operatorname{Im}\omega_{\boldsymbol k}\propto(|\Omega_{\boldsymbol k}|^2-\chi^2\Delta_{\boldsymbol k}^2)^{1/2}<\gamma_{\boldsymbol 0}$, so modes near the cutoff $k_c$ accumulate little population by the squeezing-minimum time $t_{\min}\sim\ln(1/\xi^2_{\min})/\Gamma_0$.
For power-law pairing, $|\Omega_{\boldsymbol k}|$ also decreases from $\Omega_0$ with increasing $|\boldsymbol k|$, further slowing the counted modes and helping explain the smaller prefactors $C$ and exponents $q$ at equal $f_m$.
Even so, $f_m$ is a simple control parameter across models, dimensions, and interaction forms; we test its predictive power in Sec.~\ref{sec:composition} by scaling $\chi$ with system size.

\section{Engineering the system-size scaling of squeezing}
\label{sec:composition}

\subsection{Protection and squeezing scaling}
We consider scaling the protection strength with linear size according to
\begin{equation}
 \chi(L)\propto L^s.
 \label{eq:protection_path}
\end{equation}
Away from the marginal case, combining the unstable-fraction scaling with Eq.~\eqref{eq:empirical_composition} gives
\begin{align}
 f_m&\propto L^{-ds/\nu},
 &m&\propto L^{d(1-s/\nu)},\nonumber\\
 \xi^2_{\min}&\propto L^{-dq s/\nu}
 \propto N^{-u},
 &u&=q\frac{s}{\nu}.
 \label{eq:composition}
\end{align}
Equation~\eqref{eq:composition} therefore predicts a family of partially collective regimes whose squeezing exponent $u=qs/\nu$ is continuously tunable via $s$.
The overall prefactor $C$ is required to predict the absolute value of optimal squeezing and will be used in the preparation-time analysis in Sec.~\ref{sec:time}.

At fixed gap protection strength ($s=0$) the unstable fraction remains finite, and Eq.~\eqref{eq:empirical_composition} predicts nonscalable, size-independent squeezing.
For $0<s<\nu$, squeezing improves with size even though the number of unstable modes grows, because their growth is subextensive and their fraction decreases.
Away from the marginal case $\alpha=d+2$, setting $s=\nu$ keeps the number of unstable modes of order one, though a finite number of unstable modes alone does not guarantee Heisenberg scaling.
For $s=\nu$, $\chi\Delta_{\boldsymbol k}$ at fixed $n_\mu=k_\mu L/2\pi$ is independent of $L$, so the growth rates of the remaining unstable modes relative to the collective one stay fixed.

In one dimension at $\alpha=3$, the logarithm in the protection dispersion makes the unstable fraction decrease faster with $L$ than the power law $L^{-s/2}$.
The logarithm enhances the scaling of squeezing relative to the leading power with a correction that decreases as system size grows, as derived in Appendix~\ref{app:composition_calibration}.

\subsection{Engineering the partially collective phase}
We test Eq.~\eqref{eq:composition} directly by scaling the protection as $\chi\propto L^s$ with $0\le s/\nu\le1$ in one dimension ($\alpha=1.5,2,3,4$) and $s/\nu=1/4,1/2,3/4$ on the square lattice ($\alpha=3$), for both pairing interactions.
Along each protection path, we extract $u_{\mathrm{obs}}$ from the size dependence of the dTWA minima, $\xi^2_{\min}\propto N^{-u_{\mathrm{obs}}}$, and compare it with $u=qs/\nu$, taking $q$ from the fixed-size scans of Sec.~\ref{sec:mode_fraction}.
The prediction therefore involves no fit to the size dependence.
The protection paths are specified in Appendix~\ref{app:composition_calibration}.

\begin{figure}[tbp]
 \centering
 \includegraphics[width=\columnwidth]{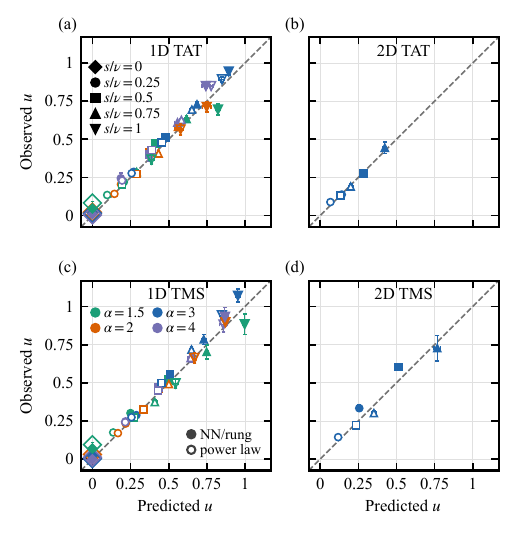}
\caption{Engineering the scaling exponent $u$ of squeezing defined by $\xi^2_{\min}\propto N^{-u}$.
Panels (a) and (b) show TAT in one and two dimensions, respectively; (c) and (d) show the corresponding TMS results.
Observed exponents extracted from dTWA simulations are compared with the prediction $u=qs/\nu$ (including the leading logarithmic correction for one-dimensional $\alpha=3$, $s>0$).
The dashed diagonal line indicates equality between prediction and simulation.
Vertical error bars show standard errors of the fitted size-scaling exponents, and horizontal error bars reflect the fitting uncertainty in $q$.}
 \label{fig:composition}
\end{figure}

Figure~\ref{fig:composition} compares the observed squeezing exponents with the predictions of Eq.~\eqref{eq:composition}.
For one-dimensional $\alpha=3$ and $s>0$, the predictions include the leading logarithmic correction of Eq.~\eqref{eq:composition_log} (Appendix~\ref{app:composition_calibration}), evaluated over the same system sizes as the observed exponent.
In one dimension, the observed exponents increase with $s$ for all considered models.
The leading logarithmic term accounts for much of the enhancement above $qs/2$ at $\alpha=3$, bringing the agreement to the level found for the other values of $\alpha$.
The remaining departures are consistent with corrections to the collapse onto $f_m$ and to the single power law of Eq.~\eqref{eq:empirical_composition}.

The square-lattice results confirm that the same scaling holds in two dimensions.
Both local and power-law pairing show the expected increase of $u$ with $s/\nu$, and the predicted and observed exponents agree across models and values of $\alpha$.
Over the accessible lattice sizes, two-dimensional rung TMS exhibits the largest finite-size departures.

\section{Preparation time of squeezed states}
\label{sec:time}
Stronger gap protection yields deeper squeezing minima.
However, when both the squeezing and gap-protecting terms are engineered from the same native interaction, or applied sequentially in Trotterized schemes, they share a common resource, so stronger protection slows the squeezing dynamics.
If the total preparation time is also limited, for example by decoherence, stronger protection is not always optimal: weaker protection can yield stronger squeezing within the available time~\cite{Kaubruegger2019}.

We now examine this competition for the four one-dimensional models with dipolar protection, $\alpha=3$, using $L=200,400,800,1600$.
We focus specifically on dipolar interactions because they are native to polar molecules and Rydberg atoms, where Floquet engineering has realized two-axis-twisting mean-field dynamics~\cite{Miller2024} and the relative strength of Heisenberg exchange to squeezing terms can be tuned~\cite{koyluoglu2025squeezingheisenberglimitlocally}.
We determine the protection strength that minimizes the squeezing parameter at fixed preparation time.

\subsection{Squeezing dynamics and preparation time}

To predict squeezing at a finite preparation time, we combine the minimum-squeezing law of Eq.~\eqref{eq:empirical_composition} with the collective dynamics derived in Appendix~\ref{app:v:continuous}.
Initially, $\xi^2\simeq e^{-\Gamma_0t}$, with the collective rates $\Gamma_0$ from Sec.~\ref{sec:bogoliubov}.
Finite spin length and the decreasing polarization modify this exponential evolution as the squeezing develops.

The unstable noncollective modes limit squeezing as if only a reduced number of spins squeezed collectively.
We make this explicit by modeling the time evolution with the finite-size matched collective trajectory $G_{\mathcal M}(\tau;M_{\mathrm{eff}})$ derived in Appendix~\ref{app:v:continuous}, where $\tau=\Gamma_0 t$ and $\mathcal M\in\{\mathrm{TAT},\mathrm{TMS}\}$, choosing the continuous effective collective size $M_{\mathrm{eff}}$ so that its first minimum equals the independently fitted value $\widehat{\xi^2}_{\min}=Cf_m^q$.
In the large-$M$ collective limit, the minimum is $\xi^2_{\min}\simeq F_{\mathcal M}(z_*)/M$ (with $M$ denoting total spins for TAT and spins per layer for TMS), where $F_{\mathrm{TAT}}(z_*)\simeq3.91$ and $F_{\mathrm{TMS}}(z_*)\simeq2.66$ (Appendices~\ref{app:v:tmscollective} and~\ref{app:v:tatcollective}).
To leading order in $1/M_{\mathrm{eff}}$, this gives
\begin{equation}
 M_{\mathrm{eff}}\simeq\frac{F_{\mathcal M}(z_*)}{C f_m^q},
 \label{eq:v:meff}
\end{equation}
which is independent of $L$ at fixed $f_m$ and scales as $L^d/m$ for $q\simeq1$.
The matched trajectory $\widehat{\xi^2}(\tau)=G_{\mathcal M}(\tau;M_{\mathrm{eff}})$ then predicts the full time dependence up to its first minimum $\tau_B=\Gamma_0\widehat t_{\min}$ without further parameters.
Writing $\ell=-\ln\widehat{\xi^2}_{\min}$, strong squeezing gives
\begin{equation}
 \Gamma_0\widehat t_{\min}\simeq\ell+\Delta_{\mathrm{coll}},
 \qquad
 \Delta_{\mathrm{coll}}^{\mathrm{TAT}}\simeq1.04,\quad
 \Delta_{\mathrm{coll}}^{\mathrm{TMS}}\simeq0.97,
 \label{eq:v:clock}
\end{equation}
i.e., the logarithmic time of exponential squeezing plus an $O(1)$ delay due to depletion of the polarized state.
All plotted predictions and Table~\ref{tab:lab_times} evaluate $M_{\mathrm{eff}}$ and $\tau_B$ numerically from the full matched trajectory $G_{\mathcal M}(\tau;M_{\mathrm{eff}})$, which retains finite-$M_{\mathrm{eff}}$ corrections to Eqs.~\eqref{eq:v:meff} and~\eqref{eq:v:clock}.

To account for engineering the protection and squeezing terms from a common interaction resource, we assume that their spin components are applied sequentially, giving a time overhead $D=3\chi+2$.
For reference interaction strength $J_{\mathrm{ref}}$ and laboratory time $T_{\mathrm{lab}}$,
\begin{equation}
 \mathcal T=J_{\mathrm{ref}}T_{\mathrm{lab}}=(3\chi+2)t.
 \label{eq:v:labclock}
\end{equation}
Increasing $\chi$ therefore reduces the initial physical squeezing rate to $J_{\mathrm{ref}}\Gamma_0/(3\chi+2)$.
Other implementations with $D\propto\chi$ at large $\chi$ give the same leading preparation-time scaling (Appendix~\ref{app:v:budgets}).

\begin{figure}[tbp]
 \centering
 \includegraphics[width=\columnwidth]{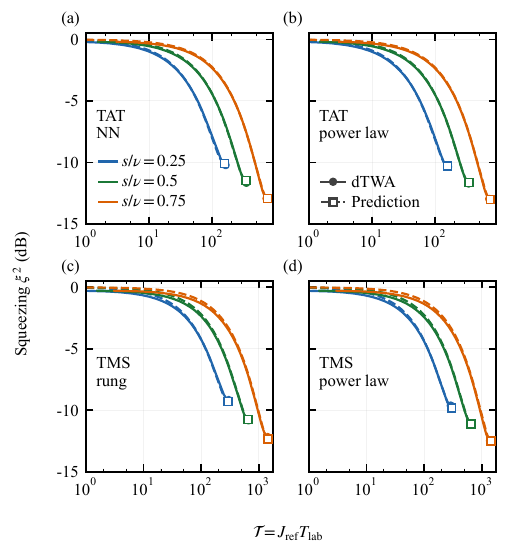}
\caption{Time evolution of squeezing at $L=1600$ and $\alpha=3$ for three choices of $s/\nu$, with $\mathcal T=J_{\mathrm{ref}}T_{\mathrm{lab}}=(3\chi+2)t$.
Solid curves show dTWA and dashed curves show the matched analytical prediction of Appendix~\ref{app:v:continuous}, whose minimum is fixed by $C f_m^q$.
Filled circles and open squares mark the dTWA and predicted minima, respectively.
Weaker protection reaches moderate squeezing sooner, whereas stronger protection reaches a deeper minimum later.}
 \label{fig:v:trajectories}
\end{figure}

Figure~\ref{fig:v:trajectories} compares the predicted trajectories with dTWA at $L=1600$, and the corresponding minimum times are benchmarked in Fig.~\ref{fig:v:timeparity} (Appendix~\ref{app:v:continuous}).

\subsection{Optimal protection at fixed preparation time}

At fixed system size $L$ and available preparation time $\mathcal T$, we seek the protection strength $\chi$ that minimizes the squeezing parameter.
For each $\chi$, the system evolves for the available time $\Gamma_0\mathcal T/(3\chi+2)$ or is stopped at its first minimum $\tau_B$, whichever comes first, giving
\begin{equation}
 \widehat{\xi^2}_{\mathrm{budget}}(L,\chi;\mathcal T)
 =\widehat{\xi^2}\!\left(\min\!\left[\frac{\Gamma_0\mathcal T}{3\chi+2},\,\tau_B\right]\right).
 \label{eq:v:budget}
\end{equation}
Minimizing Eq.~\eqref{eq:v:budget} over $\chi$ within the simulated range of protection strengths (Appendix~\ref{app:v:continuous}) gives the optimal squeezing $\widehat{\xi^2}_{\mathrm{opt}}(\mathcal T;L)$ and protection $\chi_\star(\mathcal T;L)$.

\begin{figure*}[tbp]
 \centering
 \includegraphics[width=\textwidth]{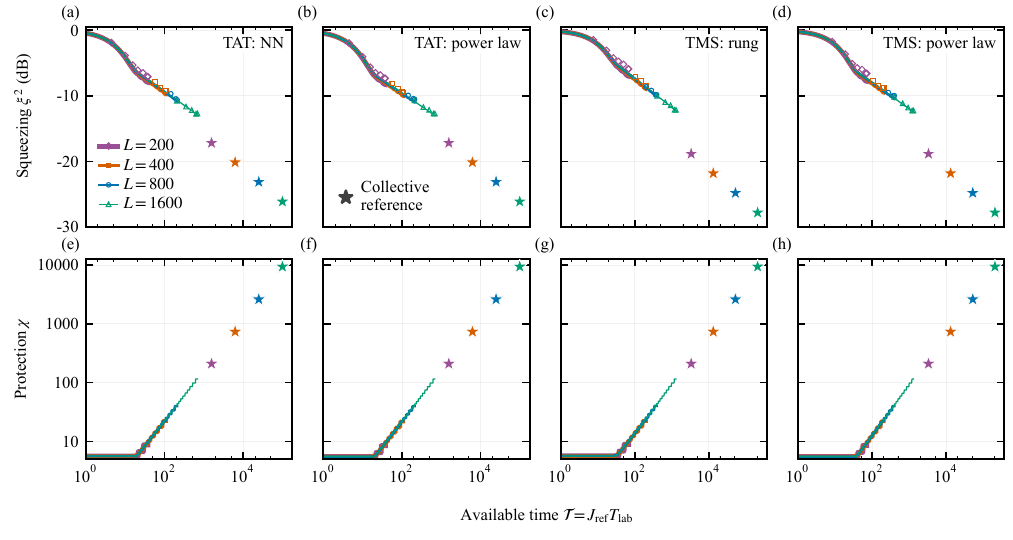}
\caption{Optimal Wineland squeezing $\xi^2$ and protection strength $\chi$ at $\alpha=3$ versus total available preparation time $\mathcal T=J_{\mathrm{ref}}T_{\mathrm{lab}}=(3\chi+2)t$.
(a)--(d) Best squeezing within the available time for $L=200,400,800,1600$.
Open markers show dTWA at the predicted optimal protection $\chi_\star$ and stopping time, wherever $\chi_\star$ coincides with a simulated protection strength.
(e)--(h) Corresponding optimal protection strength.
Each curve ends when the lowest predicted squeezing within the sampled protection range is first reached.
Stars show the exact finite-$L$ ideal collective minimum in (a)--(d) and the stability threshold $\chi_c(L)$ in (e)--(h), at the reference time defined in Appendix~\ref{app:v:collective_reference}.}
 \label{fig:v:exactL}
\end{figure*}

Figure~\ref{fig:v:exactL} shows the optimal squeezing (upper row) and corresponding protection strength (lower row), obtained by minimizing Eq.~\eqref{eq:v:budget} over $\chi$.
The corresponding $\alpha=2$ results are presented in Appendix~\ref{app:v:alpha2}.
As the available preparation time increases, the optimal protection shifts from weak $\chi$, which squeezes fastest, to stronger $\chi$, which permits deeper squeezing, until the strongest available protection reaches its minimum and longer times yield no further improvement.
The discrete changes of $m$ with $\chi$ produce the steps in the optimal protection.
The stars in Fig.~\ref{fig:v:exactL} place these preparations alongside ideal collective squeezing at the resource cost of the stability threshold $\chi_c(L)$, with time coordinate combining the exact finite-$L$ collective minimum time and $D[\chi_c(L)]$ (Appendix~\ref{app:v:collective_reference}).

dTWA simulations at the predicted optimal protection and stopping time (open markers in Fig.~\ref{fig:v:exactL}) confirm these predictions in the full dynamics.
This trade-off is evident in the dTWA trajectories of Fig.~\ref{fig:v:trajectories}: among curves reaching a given squeezing level, weaker protection arrives first.

The preparation time also sets the cost of scalable squeezing.
Along a protection path $\chi\propto L^s$, the time to the squeezing minimum grows only logarithmically with $L$ [Eq.~\eqref{eq:v:clock}], whereas the overhead $D\propto\chi$ grows as $L^s$, so that $\mathcal T_{\min}\propto L^s\ln L$ for $s>0$ (Appendix~\ref{app:composition_calibration}).
Thus stronger improvement of squeezing with size also increases the preparation cost.
In particular, suppressing all finite-momentum instabilities ($s=\nu$) costs $\mathcal T_{\min}\propto N^{\nu/d}\ln N$ (up to logarithmic corrections at $\alpha=d+2$).
This cost is superlinear for one-dimensional dipolar chains ($\nu/d=2$), whereas higher-dimensional geometries with $\nu<d$ admit sublinear collective preparation times~\cite{koyluoglu2025squeezingheisenberglimitlocally}, consistent with power-law speed limits for entanglement generation~\cite{Tran2021,Chen2023RPP}.

The same relations determine how the best squeezing improves with the available time.
Away from the marginal exponent, $f_m\propto\chi^{-d/\nu}$ gives $\ell=-\ln\widehat{\xi^2}_{\min}=(qd/\nu)\ln\chi+\mathrm{const.}$
For $\chi\gg1$, $D\simeq3\chi$, and the optimal protection approximately satisfies $\Gamma_0\mathcal T\simeq D(\chi_\star)\,\tau_B(\chi_\star)$.
To leading logarithmic order, this gives
\begin{equation}
 \chi_\star\propto\frac{\Gamma_0\mathcal T}{\ln(\Gamma_0\mathcal T)},\qquad
 \widehat{\xi^2}_{\mathrm{opt}}\propto\left[\frac{\Gamma_0\mathcal T}{\ln(\Gamma_0\mathcal T)}\right]^{-qd/\nu}.
 \label{eq:v:bestscaling}
\end{equation}
Equivalently, reaching a target $\xi^2_\star$ requires $\mathcal T\propto(\xi^2_\star)^{-\nu/(qd)}\ln(1/\xi^2_\star)$.
When protection and squeezing share an interaction resource, deeper squeezing therefore requires a preparation time that grows as a power of $1/\xi^2_\star$, rather than logarithmically as in all-to-all collective TMS and TAT.
At the marginal case $d=1$ and $\alpha=3$, the logarithmic factors cancel between $f_m\propto(\chi\ln\chi)^{-1/2}$ and $\Gamma_0\mathcal T\propto\chi\ln\chi$, giving the pure power law $\widehat{\xi^2}_{\mathrm{opt}}\propto(\Gamma_0\mathcal T)^{-q/2}$.
By comparison, the ideal collective-threshold stars in Fig.~\ref{fig:v:exactL} follow $(\xi^2_{\min})^{\mathrm{coll}}\propto(\mathcal T_{\mathrm{ref}}/\ln\mathcal T_{\mathrm{ref}})^{-d/\nu}$ away from $\alpha=d+2$, and $(\xi^2_{\min})^{\mathrm{coll}}\propto \mathcal T_{\mathrm{ref}}^{-1/2}$ at $d=1$ and $\alpha=3$ (where the logarithms cancel between $\chi_c\propto L^2/\ln L$ and $\tau_{\min}\sim\ln L$).
Because $q<1$ across most models, the optimal partially collective squeezing at fixed preparation time scales with a slightly shallower algebraic exponent.

\subsection{Implementation with dipolar spin exchange}
\label{sec:implementation}

The TAT model with power-law pairing can be realized directly from native dipolar spin exchange,
\begin{equation}
 \hat H_{\mathrm{nat}}=J_{\mathrm{nn}}\sum_{i<j}W_{ij}\,(\hat s_i^x\hat s_j^x+\hat s_i^y\hat s_j^y),
 \qquad W_{ij}=r_{ij}^{-3},
 \label{eq:v:native}
\end{equation}
as available with Rydberg atoms~\cite{Bornet2023a,Scholl2022} and polar molecules~\cite{Miller2024,holland2026creatingprobingspinsqueezedstates} for a quantization axis perpendicular to the chain.
Global microwave $\pi/2$ pulses rotate the exchange plane among the $yz$, $zx$, and $xy$ planes~\cite{Scholl2022,Miller2024}.
Spending the fractions
\begin{equation}
 w_x=\frac13,\qquad w_y=\frac{\chi-2v}{3\chi},\qquad w_z=\frac{\chi+2v}{3\chi}
 \label{eq:v:fractions}
\end{equation}
of each Floquet cycle with the exchange in the $yz$, $zx$, and $xy$ planes, respectively, yields the average Hamiltonian
\begin{equation}
 \hat H_{\mathrm{eff}}=-\frac{J_{\mathrm{nn}}}{3\chi}\,\hat H_{\mathrm{TAT}},
 \label{eq:v:heff}
\end{equation}
with $\hat H_{\mathrm{TAT}}$ given by Eq.~\eqref{eq:tat_hamiltonian} and pairing strength $v=1/\sum_jW_{ij}\simeq1/[2\zeta(3)]\approx0.42$ setting $V_{ij}=vW_{ij}$.
For the polarized initial state considered here, reversing the overall Hamiltonian sign leaves the squeezing parameter unchanged.
Nonnegative cycle fractions require $\chi\ge2v\approx0.83$, equivalently $v/\chi\le1/2$, as obtained for native XY interactions in Ref.~\cite{koyluoglu2025squeezingheisenberglimitlocally}.
The interaction ratio realized in Ref.~\cite{Miller2024} corresponds to $\chi=2v$, up to a relabeling of spin axes.
Equation~\eqref{eq:v:heff} gives $\mathcal T=|J_{\mathrm{nn}}|T_{\mathrm{lab}}=3\chi t$, i.e., $D=3\chi$, compared with the sequential-component estimate $D=3\chi+2$ used above.

For $^{87}$Rb atoms in the $60S_{1/2}$ and $60P_{3/2}$ Rydberg states at lattice spacing $a=15\,\mu\mathrm{m}$, $|J_{\mathrm{nn}}|=2\pi\times0.5\,\mathrm{MHz}$~\cite{Bornet2023a}, and $n=75$ states at $a=19\,\mu\mathrm{m}$ give $|J_{\mathrm{nn}}|\approx2\pi\times1.1\,\mathrm{MHz}$~\cite{Scholl2022}.
Microwave Floquet sequences with cycle times of $\approx0.3\,\mu\mathrm{s}$ have been applied for up to a few microseconds~\cite{Bornet2023a,Scholl2022}, while the effective lifetimes of these states at $300\,\mathrm{K}$, $100$--$220\,\mu\mathrm{s}$~\cite{Beterov2009}, limit the available time.
CaF molecules in a one-dimensional tweezer array of eight molecules with spacing $2.1\,\mu\mathrm{m}$ have $|J_{\mathrm{nn}}|=2\pi\times33\,\mathrm{Hz}$, Floquet periods of $0.5$--$0.8\,\mathrm{ms}$, and coherence times $T_2\approx0.4$--$1\,\mathrm{s}$ of the interacting rotational qubit under these sequences~\cite{holland2026creatingprobingspinsqueezedstates}.

\begin{table}[tbp]
\caption{Ideal preparation-time estimates for the one-dimensional dipolar TAT model with power-law pairing ($\alpha=3$), implemented with $D=3\chi$.
Laboratory times use $|J_{\mathrm{nn}}|=2\pi\times0.5\,\mathrm{MHz}$ for Rydberg atoms and $2\pi\times33\,\mathrm{Hz}$ for CaF molecules.
Unmarked entries use the matched analytical prediction at the optimal protection $\chi=\chi_\star$.
Daggers denote exact collective references assigned the threshold protection cost $D=3\chi_c(N)$ (showing both intermediate squeezing targets and the collective minimum for each $N$, analogous to the stars in Fig.~\ref{fig:v:exactL} which use $D=3\chi_c+2$).}
\label{tab:lab_times}
\begin{ruledtabular}
\begin{tabular}{rcrrrr}
$N$ & $\xi^2$ (dB) & $\chi$ & $\mathcal T$
& Rydberg ($\mu$s) & CaF (s) \\
\hline
20  & $-7.90^\dagger$  & 3.59 & 13.8 & 4.39 & 0.0665 \\
50  & $-8$             & 5.12 & 16.6 & 5.28 & 0.0800 \\
50  & $-10^\dagger$    & 18.0 & 69.8 & 22.2 & 0.336 \\
50  & $-11.41^\dagger$ & 18.0 & 94.5 & 30.1 & 0.456 \\
100 & $-8$             & 8.50 & 27.5 & 8.77 & 0.133 \\
100 & $-10$            & 17.4 & 68.1 & 21.7 & 0.328 \\
100 & $-12^\dagger$    & 60.8 & 273  & 87.0 & 1.32 \\
100 & $-14.25^\dagger$ & 60.8 & 385  & 123  & 1.86 \\
200 & $-8$             & 8.47 & 30.1 & 9.58 & 0.145 \\
200 & $-10$            & 17.3 & 81.1 & 25.8 & 0.391 \\
200 & $-12$            & 59.3 & 271  & 86.3 & 1.31 \\
200 & $-17.18^\dagger$ & 209  & 1550 & 493  & 7.48 \\
\end{tabular}
\end{ruledtabular}
\end{table}

Table~\ref{tab:lab_times} extrapolates the matched analytical prediction to $N=50$ and $100$ (alongside $N=200$) and compares it with ideal collective references assigned the stability-threshold cost.
For $N=50$--$200$, $-8\,\mathrm{dB}$ of squeezing is predicted at intermediate protection $\chi_\star\approx5$--$9$, requiring $5$--$10\,\mu\mathrm{s}$ at the quoted Rydberg coupling strength, whereas the ideal collective reference for $N=100$ requires $123\,\mu\mathrm{s}$, comparable to single-atom decay times.
Single-particle decay imposes a stricter condition than $T_{\mathrm{lab}}\lesssim\tau_{\mathrm{dec}}$: if a fraction $\epsilon=1-e^{-T_{\mathrm{lab}}/\tau_{\mathrm{dec}}}$ of the spins decays and drops out, the squeezing of the remaining $(1-\epsilon)N$ spins degrades as $\xi^2\to(1-\epsilon)\xi^2+\epsilon/c^2$, where $c$ is the contrast (and phase sensitivity relative to the initial $N$ spins degrades further by $1/(1-\epsilon)$).
For $\tau_{\mathrm{dec}}\approx100$--$130\,\mu\mathrm{s}$~\cite{Beterov2009} and taking $c\simeq1$ as an upper bound, the $-8\,\mathrm{dB}$ Rydberg entries retain about $-6.3$ to $-7.2\,\mathrm{dB}$, whereas the collective reference for $N=100$ retains at most $-1.5$ to $-2\,\mathrm{dB}$ (and lies above the initial-$N$ standard quantum limit).
At the molecular coupling strength, the collective reference time increases from $0.46\,\mathrm{s}$ for $N=50$ to $1.86\,\mathrm{s}$ for $N=100$, approaching or exceeding the reported coherence times.
The table entries assume ideal unitary dynamics, attainable squeezing may be further degraded by decoherence, pulse errors, and positional disorder.

\section{Summary}
\label{sec:outlook}

We have studied how Heisenberg protection controls squeezing and its scaling in single-layer TAT and bilayer TMS models with local or power-law squeezing interactions.
Bogoliubov theory relates the relative strength of the gap protection term and the low-momentum behavior of the interactions to the number of unstable modes.
Our dTWA results suggest a collapse of the achievable squeezing in a given system as a function of the fraction of unstable finite-momentum modes.
At a fixed unstable-mode fraction (such as constant $\chi$), squeezing does not scale with system size; conversely, scalable squeezing persists even when the number of unstable modes grows with $N$, provided their fraction decreases.
Scaling $\chi$ with system size thus produces a family of partially collective models with a continuously tunable squeezing exponent, in quantitative agreement with our dTWA simulations across models and interaction forms.

We also analyzed the trade-off between stronger protection, which reaches deeper squeezing at later times, and weaker protection, which squeezes faster.
At fixed available preparation time, the optimal protection depends on the time budget and generally lies below the fully protected threshold, so the achievable squeezing improves algebraically, $\widehat{\xi^2}_{\mathrm{opt}}\propto(\mathcal T/\ln\mathcal T)^{-qd/\nu}$ (or $\mathcal T^{-q/2}$ at $d=1$ and $\alpha=3$), rather than exponentially as in all-to-all models.

Controlling the fraction of unstable modes thus governs both the system-size scaling of squeezing and the trade-off between squeezing depth and preparation time across lattice dimensions and interaction geometries in pair-generating spin models.

\begin{acknowledgments}
This material is based upon work supported by the Air Force Office of Scientific Research under Award No. FA9550-25-1-0340.
T.B. and L.C. acknowledge support from the National Science Foundation through NRT NSF Grant No. DGE-2510202.
This research was supported in part by grant NSF PHY-2309135 to the Kavli Institute for Theoretical Physics (KITP).
The computing for this project was performed at the High Performance Computing Center at Oklahoma State University, supported in part through the National Science Foundation grant OAC-1531128.
\end{acknowledgments}

\bibliography{main}

\appendix
\section{Numerical benchmarks}

\subsection{Statistical convergence of the dTWA simulations}
\label{app:dtwa_sampling}

\begin{figure}[tbp]
    \centering
    \includegraphics[width=\columnwidth]{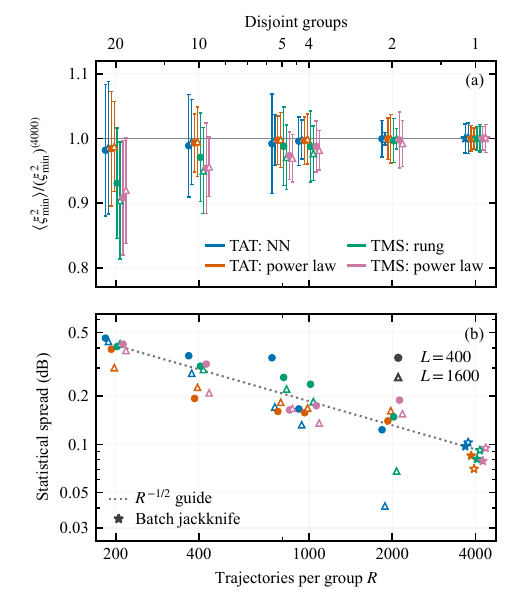}
\caption{Convergence of squeezing with trajectory number in dTWA simulations for the four one-dimensional models at $\alpha=3$, $L=400,1600$, and $s/\nu=0.75$. (a) Mean minimum squeezing over disjoint groups, normalized by the 4000-trajectory result.
Error bars show one sample standard deviation, and the top axis shows the number of distinct groups.
(b) Standard deviation of the group minima in decibels.
Stars at $R=4000$ in both panels show the batch-jackknife estimates defined in the text.
Filled circles and open triangles denote $L=400$ and $1600$, respectively.
The dotted line is an $R^{-1/2}$ guide.}
    \label{fig:dtwa_sampling}
\end{figure}

We estimate the finite-trajectory sampling uncertainty of the minimum squeezing for the four one-dimensional models at $\alpha=3$, $L=400$ and $1600$, and $s/\nu=0.75$.
We generate 20 independently seeded batches of 200 trajectories (4000 in total) and partition them into disjoint groups of $R=200$--$2000$ trajectories.
For each group, the first and second moments are pooled before the squeezing parameter is constructed and minimized in time.
Figure~\ref{fig:dtwa_sampling} shows the mean and spread of the group minima, normalized by the pooled 4000-trajectory result, together with a delete-one-batch jackknife estimate of the uncertainty of the pooled result.
With $R=200$, the minima lie $1.3$--$1.8\%$ below the pooled reference for TAT and $6.9$--$9.6\%$ below it for TMS.
At $R=2000$, the group means agree with the reference to within $0.8\%$, while individual calculations scatter by $1$--$4\%$.
The jackknife uncertainty of the pooled result is $1.6$--$2.4\%$, or $0.07$--$0.10\,\mathrm{dB}$.
These estimates quantify sampling error only; the accuracy of the dTWA approximation itself is assessed in Appendix~\ref{app:mps}.

\subsection{Matrix-product-state benchmarks}
\label{app:mps}

We benchmark dTWA against matrix-product-state (MPS) evolution for the one-dimensional models with $\alpha=3$.
The comparisons address two questions: how accurately dTWA predicts the squeezing minimum, and whether the quantum dynamics exhibit size-independent squeezing at fixed unstable-mode fraction.
Both methods use identical periodic interaction kernels and initial product states.

The MPS states are evolved using the two-site time-dependent variational principle~\cite{Haegeman2016,Paeckel2019}, with time step $\delta t=0.04$, evolution cutoff $10^{-10}$, and Hamiltonian matrix-product-operator cutoff $10^{-12}$.
We retain the largest available bond dimension, $D_{\mathrm{bond}}=256$ or $512$, and find that increasing $D_{\mathrm{bond}}$ changes the plotted minima by less than $0.04\,\mathrm{dB}$, indicating convergence over the tested range.
For the dTWA runs we again use 2000 trajectories.

\begin{figure}[tbp]
 \centering
 \includegraphics[width=\columnwidth]{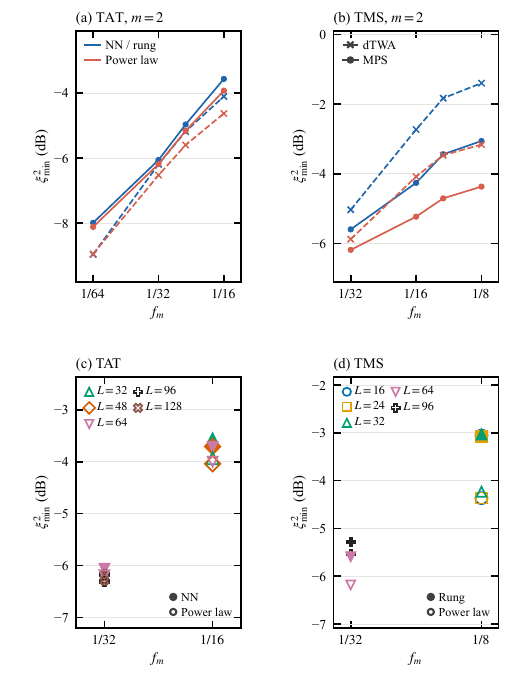}
\caption{MPS benchmarks for the one-dimensional models at $\alpha=3$, with TAT in the left column and TMS in the right column (both showing $\xi^2_{\min}$ in decibels).
(a), (b) Squeezing minima along fixed-$m=2$ paths ($L=128,64,48,32$ for TAT and $L=64,32,24,16$ for TMS, from left to right).
Blue and red distinguish local and power-law pairing.
Solid lines with circles show MPS, and dashed lines with crosses show dTWA.
(c), (d) MPS minima at fractions shared by at least two system sizes.
Colors and marker shapes identify $L$; filled and open symbols denote local and power-law pairing, respectively.}
 \label{fig:mps_benchmark}
\end{figure}

The upper panels of Fig.~\ref{fig:mps_benchmark} compare the methods along fixed-$m=2$ paths ($L=128,64,48,32$ for TAT and $L=64,32,24,16$ for TMS).
Increasing $L$ reduces $f_m=2/L$ while retaining two unstable nonzero momentum pairs.
TAT shows relatively close agreement over much of the plotted range, with dTWA slightly overestimating squeezing at the smallest $f_m$.
For TMS, MPS predicts stronger squeezing than dTWA, and the offset generally decreases toward $f_m=1/32$.

The lower panels test size dependence more directly.
At fixed $f_m$, increasing $L$ also increases the number of unstable momentum pairs.
Nevertheless, both TAT and TMS minima vary little across the available sizes at fixed $f_m$.
Thus the observed dependence on the unstable-mode fraction is also present in the MPS dynamics, even where dTWA does not reproduce the absolute squeezing accurately.

\section{Fitting procedure for the squeezing scaling}
\label{app:composition_calibration}

\begin{figure*}[tbp]
 \centering
 \includegraphics[width=\textwidth]{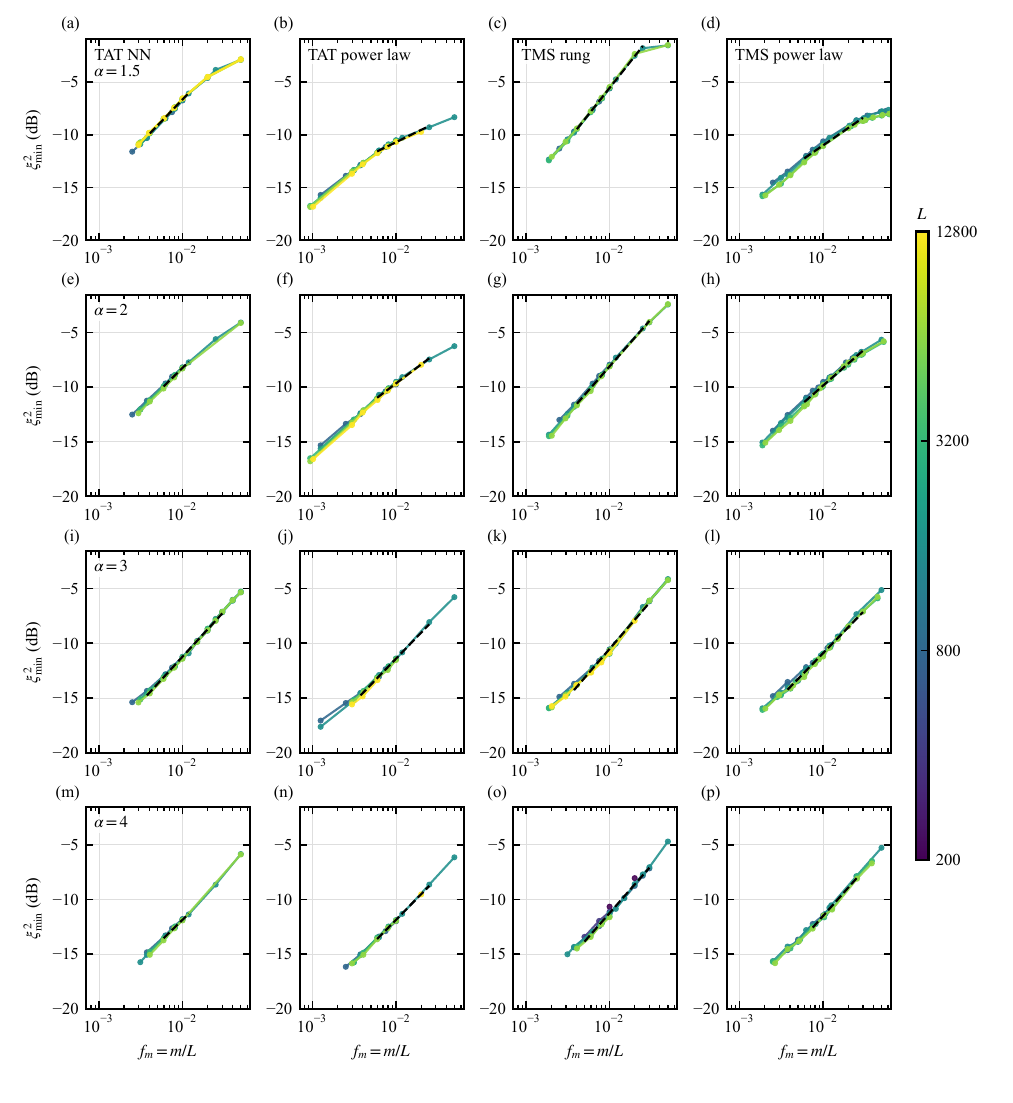}
\caption{One-dimensional squeezing scans $\xi^2_{\min}$ for local and power-law pairing at the indicated interaction exponents.
Columns distinguish the four models; rows distinguish $\alpha$.
Colors indicate $L$, and lines connect fixed-size scans.
Dashed curves show the $C f_m^q$ law over its fitted mode-fraction interval.}
 \label{fig:composition_scans_1d}
\end{figure*}

Figure~\ref{fig:composition_scans_1d} extends the one-dimensional squeezing scans to $\alpha=1.5,2,3,4$, showing good collapse onto $f_m$ for all four models.

A single power law describes a restricted interval of these curves rather than their entire range.
For each dimension, model, interaction form, and $\alpha$, we obtain $C$ and $q$ by an unweighted least-squares fit of $\ln\xi^2_{\min}$ on $\ln f_m$.
The fit uses fixed-size scan points whose mode fractions lie within the range covered by the $s/\nu=1/4,1/2,3/4$ protection paths, excluding the parameter sets on the paths themselves; Table~\ref{tab:composition_coefficients} lists the resulting fitted ranges.

\begin{table*}[tbp]
\caption{Coefficients of the squeezing law in Eq.~\eqref{eq:empirical_composition}, fitted over the indicated range of mode fractions.
Parentheses give one regression standard error in the last quoted digits.}
\label{tab:composition_coefficients}
\begin{ruledtabular}
\begin{tabular}{clcccc}
$d$ & Model/pairing & $\alpha$ &  Fitted range in $f_m$ & $C$ & $q$ \\
\hline
1 & TAT--NN & 1.5 & $[0.0041, 0.012]$ & $9.57(67)$ & $0.823(14)$ \\
1 & TAT--NN & 2 & $[0.0059, 0.012]$ & $4.72(61)$ & $0.749(27)$ \\
1 & TAT--NN & 3 & $[0.0037, 0.030]$ & $3.46(14)$ & $0.831(9)$ \\
1 & TAT--NN & 4 & $[0.0059, 0.012]$ & $2.00(26)$ & $0.744(27)$ \\
1 & TAT--power law & 1.5 & $[0.0059, 0.025]$ & $0.506(39)$ & $0.387(16)$ \\
1 & TAT--power law & 2 & $[0.0059, 0.025]$ & $1.55(11)$ & $0.578(15)$ \\
1 & TAT--power law & 3 & $[0.0037, 0.025]$ & $2.75(22)$ & $0.790(16)$ \\
1 & TAT--power law & 4 & $[0.0059, 0.025]$ & $2.35(13)$ & $0.778(12)$ \\
1 & TMS--rung & 1.5 & $[0.0041, 0.025]$ & $27.7(14)$ & $0.999(10)$ \\
1 & TMS--rung & 2 & $[0.0041, 0.030]$ & $8.59(52)$ & $0.872(13)$ \\
1 & TMS--rung & 3 & $[0.0037, 0.030]$ & $5.21(39)$ & $0.884(16)$ \\
1 & TMS--rung & 4 & $[0.0050, 0.030]$ & $4.11(40)$ & $0.869(22)$ \\
1 & TMS--power law & 1.5 & $[0.0059, 0.030]$ & $0.980(81)$ & $0.545(19)$ \\
1 & TMS--power law & 2 & $[0.0059, 0.030]$ & $2.25(13)$ & $0.669(13)$ \\
1 & TMS--power law & 3 & $[0.0037, 0.030]$ & $3.04(21)$ & $0.786(14)$ \\
1 & TMS--power law & 4 & $[0.0075, 0.026]$ & $3.71(43)$ & $0.860(26)$ \\
\hline
2 & TAT--NN & 3 & $[0.0045, 0.016]$ & $2.13(17)$ & $0.565(16)$ \\
2 & TAT--power law & 3 & $[0.0049, 0.016]$ & $0.213(20)$ & $0.266(19)$ \\
2 & TMS--rung & 3 & $[0.0052, 0.025]$ & $42.5(54)$ & $1.023(28)$ \\
2 & TMS--power law & 3 & $[0.0058, 0.016]$ & $0.476(38)$ & $0.469(17)$ \\
\end{tabular}
\end{ruledtabular}
\end{table*}

The protection paths follow $\chi(L)=\chi(L_0)(L/L_0)^s$, with $s/\nu=0,1/4,1/2,3/4,1$ in one dimension and $s/\nu=1/4,1/2,3/4$ in two dimensions.
The reference strength $\chi(L_0)$ is fixed by $f_m=0.02$ at $L_0=400$ in one dimension and by $f_m=50/80^2$ at $L_0=80$ in two dimensions, so that all paths for a given model and $\alpha$ pass through a common reference point.

The observed exponent $u_{\mathrm{obs}}$ is obtained by fitting $\ln\xi^2_{\min}$ as a function of $\ln L$, whose slope is $-du_{\mathrm{obs}}$.
The uncertainties shown in Fig.~\ref{fig:composition} are the resulting standard regression fit errors.

For $\alpha=3$ in one dimension, the long-wavelength dispersion scales as $\Delta_k\propto k^2\ln(1/|k|)$.
At the boundary $k_c$ of the unstable region, $\chi\Delta_{k_c}=|\Omega_{k_c}|$.
Since $\Omega_{k_c}\to\Omega_0\ne0$ as $k_c\to0$, a scaling of the protection strength $\chi\propto L^s$ gives $k_c^2\ln(1/k_c)\propto L^{-s}$.
For fixed $s>0$, the leading logarithmic solution is $k_c\propto L^{-s/2}(\ln L)^{-1/2}$.

As long as the unstable momenta can be treated as a continuum, $f_m\propto k_c$.
The empirical relation $\xi^2_{\min}\propto f_m^q$ then gives
\begin{align}
 f_m &\propto L^{-s/2}(\ln L)^{-1/2}, \nonumber\\
 \xi^2_{\min} &\propto L^{-qs/2}(\ln L)^{-q/2}.
 \label{eq:composition_log}
\end{align}
Since $N\propto L$ in one dimension, the local squeezing exponent is
\begin{equation}
 u_{\mathrm{eff}}(L)
 \equiv -\frac{d\ln\xi^2_{\min}}{d\ln N}
 \simeq \frac{qs}{2}+\frac{q}{2\ln L}.
 \label{eq:composition_log_local}
\end{equation}
This leading-logarithmic approximation omits constants inside the logarithm and higher logarithmic corrections.

Combined with Eq.~\eqref{eq:v:clock} and the overhead $D=3\chi+2$ of Sec.~\ref{sec:time}, the minimum preparation time along a protection path at $\alpha=3$ is
\begin{equation}
 \widehat{\mathcal T}_{\min}\simeq\frac{D}{\Gamma_0}
 \left[\frac{qs}{2}\ln L+\frac q2\ln\ln L+c_0+\Delta_{\mathrm{coll}}\right],
 \label{eq:v:scaling}
\end{equation}
where $c_0$ is independent of $L$.
Since $D\propto\chi\propto L^s$, the logarithmic corrections modify finite-size times without changing the leading dependence $\mathcal T_{\min}\propto L^s\ln L$.

\section{Squeezing dynamics and preparation times}
\label{app:v:preparation}
We first specify the interaction resource used in Sec.~\ref{sec:time}, then derive the collective squeezing functions and their early-time corrections.
Combining these limits gives the trajectory approximation used for the spatial models.
Throughout, $\Omega_0=1$, and $L$ denotes the chain size for TAT or the number of spins per layer for TMS.

\subsection{Interaction strengths and preparation time}
\label{app:v:budgets}

We assume that each Cartesian component of the protection and pairing interactions can be applied independently at coefficient strength $J_{\mathrm{comp}}$, with its prescribed spatial profile and sign.
Their relative durations are $\chi:\chi:\chi:1:1$.
For TMS, protection acts on both layers simultaneously; for TAT, the construction includes the opposite signs of the pairing components and accounts via $\kappa_{\mathrm{TAT}}=2$ for the overall factor of two in Eq.~\eqref{eq:tat_hamiltonian}.
Writing $\kappa_{\mathrm{TAT}}=2$ and $\kappa_{\mathrm{TMS}}=1$, the short-cycle average is
\begin{equation}
 \hat H_{\mathrm{eff}}=\frac{J_{\mathrm{comp}}}{\kappa_{\mathcal M}(3\chi+2)}\hat H_{\mathcal M},
\end{equation}
for $\mathcal M\in\{\mathrm{TAT},\mathrm{TMS}\}$.
Setting $J_{\mathrm{ref}}=J_{\mathrm{comp}}/\kappa_{\mathcal M}$ recovers Eq.~\eqref{eq:v:labclock}, $\mathcal T=J_{\mathrm{ref}}T_{\mathrm{lab}}=(3\chi+2)t$.
This counts the separately applied spin components while leaving the cost of engineering the spatial profiles unspecified.
The normalization is that of the Hamiltonians in Sec.~\ref{sec:model}: the pairing row sum is one, whereas the protection row sum need not be one.

For native pairing couplings with row sum $\Omega_0^{\mathrm{nat}}$, this normalization gives $V_{ij}=V_{ij}^{\mathrm{nat}}/\Omega_0^{\mathrm{nat}}$, $\chi=\chi_{\mathrm{nat}}/\Omega_0^{\mathrm{nat}}$, and $t=\Omega_0^{\mathrm{nat}}t_{\mathrm{nat}}$.
The protection profile $W_{ij}$ retains its spatial form while its coefficient is rescaled, leaving the squeezing unchanged.

\subsection{Correction to exponential squeezing}
\label{app:v:delta}

For an exactly normalized trajectory with $\xi^2(0)=1$, define $\tau=\Gamma_0t$ and $R(\tau)=-d\ln\xi^2/d\tau$.
The deviation from ideal exponential squeezing obeys
\begin{equation}
 \Delta=\tau_{\min}+\ln\xi^2_{\min}
 =\int_0^{\tau_{\min}}[1-R(\tau)]\,d\tau.
 \label{eq:v:deltaintegral}
\end{equation}
The undepleted approximation gives $R=1$.
Finite spin length changes the pair-production matrix elements, while depolarization reduces the signal entering the Wineland parameter.
Related analyses beyond the undepleted-pump approximation have been carried out for cavity models with additional self-interactions~\cite{Sundar2023}; below we derive the corresponding expressions directly for the collective limits of our Hamiltonians, retaining both effects.

\subsection{Collective two-mode squeezing}
\label{app:v:tmscollective}

For uniform interlayer pairing, $V_{ij}=\Omega_0/L$, each layer remains in its symmetric sector $J=L/2$, with $N=2L$ spins in total.
Intralayer Heisenberg exchange is constant in these sectors, and
\begin{equation}
 \hat H=\frac{\Omega_0}{2L}(\hat S_A^-\hat S_B^++\hat S_A^+\hat S_B^-).
 \label{eq:v:tmscollectiveH}
\end{equation}
Starting from opposite polarizations, write
\begin{gather}
 |\psi(\tau)\rangle=\sum_{n=0}^{L}(-i)^n b_n(\tau)|n\rangle,\nonumber\\
 |n\rangle=|J,J-n\rangle_A|J,-J+n\rangle_B,
 \qquad \tau=\Omega_0t.
\end{gather}
Here $n$ counts pairs, with one flipped spin in each layer.
The amplitudes $b_n$ are real, $b_n(0)=\delta_{n0}$, and $\sum_n b_n^2=1$.
Each ladder operator contributes $\sqrt{(n+1)(L-n)}$, giving
\begin{align}
 h_n&=\frac{\langle n+1|\hat H|n\rangle}{\Omega_0}
     =\frac{(n+1)(L-n)}{2L},\nonumber\\
 \partial_\tau b_n&=h_{n-1}b_{n-1}-h_nb_{n+1},
 \label{eq:v:tmsamplitudes}
\end{align}
with $h_{-1}=h_L=0$.

On the first squeezing branch, $\hat Q_-=(\hat S_A^x-\hat S_B^y)/\sqrt2$ has zero mean and minimum transverse variance.
Acting with $\hat Q_-$ takes neighboring pair sectors to the same intermediate states.
Their interference gives
\begin{align}
 \xi^2_{\mathrm{var}}
 &=\frac{4\operatorname{Var}(\hat Q_-)}{L}\nonumber\\
 &=\frac1L\sum_{n=0}^{L-1}(n+1)(L-n)(b_{n+1}-b_n)^2.
 \label{eq:v:tmspositive}
\end{align}
While the layer polarizations retain their initial signs, the contrast is $c=1-2\langle n\rangle/L$, and $\xi^2_{\mathrm{TMS}}=\xi^2_{\mathrm{var}}/c^2$.
These expressions and the amplitude equation are exact at finite $L$.

In the undepleted limit, $b_n=\operatorname{sech}(\tau/2)\tanh^n(\tau/2)$ and $\langle n\rangle=\sinh^2(\tau/2)$.
At large $L$, introduce $x=n/L$ and $b_n\simeq\sqrt{p(x,\tau)/L}$.
Expanding Eq.~\eqref{eq:v:tmsamplitudes} gives
\begin{equation}
 \partial_\tau p+\partial_x[x(1-x)p]=0.
 \label{eq:v:collectivetransport}
\end{equation}
The factor $1-x$ describes depletion of the initially polarized populations.
In the overlap regime $1\ll\langle n\rangle\ll L$, the geometric pair distribution approaches $n\simeq e^\tau Y/4$, where $Y\ge0$ has exponential probability density $e^{-Y}$.
Integrating the drift gives
\begin{gather}
 x=\frac{zY}{1+zY},\qquad z=\frac{e^\tau}{4L},\nonumber\\
 p(x)=\frac{\exp[-x/(z(1-x))]}{z(1-x)^2}.
\end{gather}
The continuum limit of Eq.~\eqref{eq:v:tmspositive} is
\begin{align}
 \xi^2_{\mathrm{var}}
 &\simeq\frac1L\int_0^1x(1-x)[\partial_x\sqrt{p(x)}]^2\,dx\nonumber\\
 &=\frac1L\left(\frac1{4z}+\frac z2\right).
 \label{eq:v:tmstransport}
\end{align}
The same distribution gives
\begin{equation}
 c_{\mathrm{TMS}}(z)=2\left\langle\frac1{1+zY}\right\rangle-1
 =\frac{2e^{1/z}}{z}E_1(1/z)-1,
\end{equation}
where $E_1(y)=\int_y^\infty e^{-v}\,dv/v$.
Therefore
\begin{equation}
 L\xi^2_{\mathrm{TMS}}\simeq F_{\mathrm{TMS}}(z)
 =\frac{1/(4z)+z/2}{c_{\mathrm{TMS}}(z)^2}.
 \label{eq:v:tmsF}
\end{equation}
On the first positive-contrast branch, the minimum occurs at $z_*\simeq0.24947$, with $F_{\mathrm{TMS}}(z_*)\simeq2.65646$.
Hence
\begin{equation}
 \tau_{\min}\simeq\ln(4Lz_*),\qquad
 \xi^2_{\min}\simeq\frac{F_{\mathrm{TMS}}(z_*)}{L},
 \label{eq:v:tmsdelta}
\end{equation}
and $\Delta_{\mathrm{coll}}^{\mathrm{TMS}}=\ln[4z_*F_{\mathrm{TMS}}(z_*)]\simeq0.97488$.
This limit keeps $\tau-\ln L$ fixed as $L\to\infty$ and retains the full pair-number distribution.

\begin{figure}[tbp]
 \centering
 \includegraphics[width=\columnwidth]{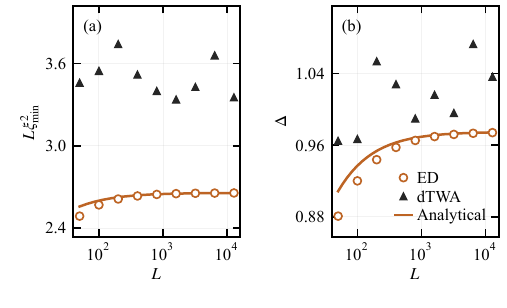}
\caption{Collective (all-to-all) TMS: (a) scaled minimum Wineland squeezing $L\xi^2_{\min}$ and (b) timing correction $\Delta=\tau_{\min}+\ln\xi^2_{\min}$.
Circles denote ED, triangles dTWA, and solid curves the finite-size matched prediction of Eq.~\eqref{eq:v:matched_derivation} (Appendix~\ref{app:v:continuous}), evaluated at $M=L$.
Here $L$ is the number of spins per layer.}
 \label{fig:v:collective_tms}
\end{figure}

Figure~\ref{fig:v:collective_tms} compares the matched analytical prediction with exact collective-spin evolution (ED) and dTWA (2000 trajectories).
The analytical results approach ED for both observables as $L$ increases, while dTWA gives larger minima.
A related comparison for the TMS variance is given in Ref.~\cite{PhysRevA.109.L061304}.

\subsection{Collective two-axis twisting}
\label{app:v:tatcollective}

For uniform pairing, the dynamics remain in the symmetric sector $J=N/2$, where Heisenberg protection is constant.
Rotating the initial polarization to $+z$ gives
\begin{equation}
 \hat H=\frac{\Omega_0}{N-1}\bigl[(\hat S^x)^2-(\hat S^y)^2\bigr],
 \qquad \tau=2\Omega_0t,\quad N=L.
 \label{eq:v:tatcollectiveH}
\end{equation}
The denominator excludes self-coupling while fixing the pairing row sum.
In the basis $|n\rangle=|J,J-n\rangle$, $n$ counts individual spin flips.
Since $(\hat S^x)^2-(\hat S^y)^2=[(\hat S^+)^2+(\hat S^-)^2]/2$, only $n\leftrightarrow n+2$ transitions occur.
Writing the amplitude as $(-i)^{n/2}b_n$ for even $n$ gives real $b_n$, with $b_n(0)=\delta_{n0}$ and
\begin{align}
 \partial_\tau b_n&=h_{n-2}b_{n-2}-h_nb_{n+2},\nonumber\\
 h_n&=\frac{\sqrt{(N-n)(N-n-1)(n+1)(n+2)}}{4(N-1)}.
 \label{eq:v:tatrecurrence}
\end{align}
Here $h_n=\langle n+2|\hat H|n\rangle/(2\Omega_0)$, with couplings outside the basis set to zero.

The squeezed quadrature on the first branch is $\hat Q_-=(\hat S^x+\hat S^y)/\sqrt2$.
Parity gives $\langle \hat Q_-\rangle=0$.
Acting with $\hat Q_-$ on occupations $n$ and $n+2$ reaches the same odd state; their interference gives the variance $V_-=\operatorname{Var}(\hat Q_-)$ as
\begin{align}
 V_-=\frac14\sum_{\substack{0\le n<N\\n\;\text{even}}}
 \bigl[&\sqrt{(N-n)(n+1)}b_n\nonumber\\
 &-\sqrt{(N-n-1)(n+2)}b_{n+2}\bigr]^2.
 \label{eq:v:tatpositive}
\end{align}
As for TMS, the Wineland parameter includes the remaining polarization: $\xi^2_{\mathrm{TAT}}=NV_-/\langle \hat S^z\rangle^2$.

For $x=n/N$, the large-$N$ probability density obeys $b_n\simeq\sqrt{2p(x)/N}$; the factor of two accounts for the spacing of allowed occupations.
Expanding Eq.~\eqref{eq:v:tatrecurrence} gives the same drift $dx/d\tau=x(1-x)$ as for TMS.
The undepleted state is a single-mode squeezed vacuum, whose amplified Gaussian quadrature gives $Y=Z^2$ for a standard normal variable $Z$.
Matching to $\langle n\rangle\simeq e^\tau/4$ gives $x=zY/(1+zY)$, with $z=e^\tau/(4N)$, and
\begin{equation}
 p(x)=\frac{x^{-1/2}(1-x)^{-3/2}}{\sqrt{2\pi z}}
 \exp\!\left[-\frac{x}{2z(1-x)}\right].
\end{equation}
Equation~\eqref{eq:v:tatpositive} then yields
\begin{align}
 V_-\simeq{}&\frac14\int_0^1x(1-x)
 \left[2\partial_x\sqrt p+
 \frac{1-2x}{2x(1-x)}\sqrt p\right]^2\,dx\nonumber\\
 ={}&\frac1{16z}+\frac18+\frac{7z}{16}.
\end{align}
Since $\hat S^z=N/2-\hat n$, the contrast is
\begin{equation}
 c_{\mathrm{TAT}}(z)=2\left\langle\frac1{1+zY}\right\rangle-1
 =2\sqrt{\frac{\pi}{2z}}\operatorname{erfcx}
 \!\left(\frac1{\sqrt{2z}}\right)-1.
\end{equation}
Here $\operatorname{erfc}(y)=(2/\sqrt\pi)\int_y^\infty e^{-v^2}\,dv$ and $\operatorname{erfcx}(y)=e^{y^2}\operatorname{erfc}(y)$.
Thus
\begin{equation}
 N\xi^2_{\mathrm{TAT}}\simeq F_{\mathrm{TAT}}(z)
 =\frac{1/(4z)+1/2+7z/4}{c_{\mathrm{TAT}}(z)^2}.
 \label{eq:v:tatF}
\end{equation}
The first positive-contrast minimum gives $z_*\simeq0.18031$ and $F_{\mathrm{TAT}}(z_*)\simeq3.91260$.
Consequently $\tau_{\min}\simeq\ln(4Nz_*)$, $\xi^2_{\min}\simeq F_{\mathrm{TAT}}(z_*)/N$, and $\Delta_{\mathrm{coll}}^{\mathrm{TAT}}=\ln[4z_*F_{\mathrm{TAT}}(z_*)]\simeq1.03744$.

\begin{figure}[tbp]
 \centering
 \includegraphics[width=\columnwidth]{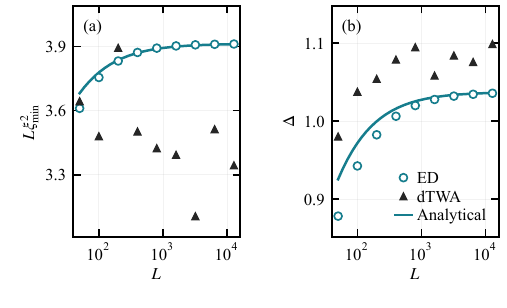}
\caption{Collective (all-to-all) TAT: (a) scaled minimum Wineland squeezing $L\xi^2_{\min}$ and (b) timing correction $\Delta=\tau_{\min}+\ln\xi^2_{\min}$.
Circles denote ED, triangles dTWA, and solid curves the finite-size matched prediction of Eq.~\eqref{eq:v:matched_derivation} (Appendix~\ref{app:v:continuous}), evaluated at $M=L$.
Here $L=N$ is the total number of spins.}
 \label{fig:v:collective_tat}
\end{figure}

Figure~\ref{fig:v:collective_tat} compares the corresponding results for TAT (again with 2000 dTWA trajectories per system size).
The matched analytical prediction approaches ED for both observables as $L$ increases, whereas dTWA gives smaller minima at the largest sizes shown.

\subsection{Time evolution from the initial state to the minimum}
\label{app:v:continuous}
The functions $F_{\mathcal M}$ describe the large-spin dynamics near the minimum.
To connect them to the initial state, we also expand the exact amplitude equations at fixed $\tau$.
Set $M=L$ for TMS and $M=N$ for TAT, and write
\begin{equation}
 b_n=b_n^{(0)}+\frac{b_n^{(1)}}{M}+O(M^{-2}).
\end{equation}
The leading amplitudes are the two-mode squeezed vacuum given above for TMS, and
\begin{equation}
 b_{2j}^{(0)}=
 \frac{\sqrt{(2j)!}}{2^jj!}
 \frac{\tanh^j(\tau/2)}{\sqrt{\cosh(\tau/2)}}
\end{equation}
for TAT.
With $\delta n=1$ for TMS and $\delta n=2$ for TAT, the matrix elements have the expansion $h_n=h_n^{(0)}[1+a_n/M+O(M^{-2})]$, where
\begin{align}
 (h_n^{(0)},a_n)_{\mathrm{TMS}}
 &=\left(\frac{n+1}{2},-n\right),\nonumber\\
 (h_n^{(0)},a_n)_{\mathrm{TAT}}
 &=\left(\frac{\sqrt{(n+1)(n+2)}}4,\frac12-n\right).
\end{align}
In particular, the TAT coefficient retains the $N-1$ normalization of Eq.~\eqref{eq:v:tatcollectiveH}.
Equating first-order terms gives the linear equation
\begin{align}
 \partial_\tau b_n^{(1)}={}&
 h_{n-\delta n}^{(0)}b_{n-\delta n}^{(1)}-h_n^{(0)}b_{n+\delta n}^{(1)}\nonumber\\
 &+a_{n-\delta n}h_{n-\delta n}^{(0)}b_{n-\delta n}^{(0)}
 -a_nh_n^{(0)}b_{n+\delta n}^{(0)},
 \label{eq:v:first_order_amplitudes}
\end{align}
with $b_n^{(1)}(0)=0$.
Substituting its solution into Eqs.~\eqref{eq:v:tmspositive} and~\eqref{eq:v:tatpositive}, and expanding their polarization denominators to the same order, gives
\begin{equation}
 \xi^2_{\mathcal M}(\tau;M)
 =e^{-\tau}+\frac{g_{\mathcal M}(\tau)}{M}+O(M^{-2}),
 \label{eq:v:early_expansion}
\end{equation}
where
\begin{align}
 g_{\mathrm{TMS}}(\tau)&=1-(1+\tau)e^{-\tau},\nonumber\\
 g_{\mathrm{TAT}}(\tau)&=\frac32-(1+2\tau)e^{-\tau}
 -\frac12e^{-2\tau}.
 \label{eq:v:early_corrections}
\end{align}
Both satisfy $g_{\mathcal M}(0)=g_{\mathcal M}'(0)=0$.
For example, the TMS variance alone has correction $e^{-\tau}(1-\tau-e^{-\tau})/M$; the remaining contribution comes from $c=1-2\sinh^2(\tau/2)/M+O(M^{-2})$.

In the common regime $1\ll e^\tau\ll M$, the early-time and near-minimum expansions both reduce to $e^{-\tau}+A_{\mathcal M}/M$, with $A_{\mathrm{TMS}}=1$ and $A_{\mathrm{TAT}}=3/2$.
Adding the two expansions and subtracting this overlap gives
\begin{align}
 G_{\mathcal M}(\tau;M)
 &=\frac{F_{\mathcal M}\!\bigl(e^\tau/(4M)\bigr)}{M}
 +\frac{g_{\mathcal M}(\tau)-A_{\mathcal M}}{M}\nonumber\\
 &=\frac{F_{\mathcal M}\!\bigl(e^\tau/(4M)\bigr)-r_{\mathcal M}(\tau)}{M},
 \label{eq:v:matched_derivation}
\end{align}
with $r_{\mathcal M}(\tau)=A_{\mathcal M}-g_{\mathcal M}(\tau)$.
This matching recovers the initial value and slope through $O(M^{-1})$.
Near $\tau\sim\ln M$, the subtraction is $O(\ln M/M^2)$, so it leaves the leading collective minimum and limiting $\Delta_{\mathrm{coll}}$ unchanged.
The residual initial normalization error is $O(M^{-2})$; it is not removed by rescaling the curves.

For the spatial models (Sec.~\ref{sec:time}), we determine the continuous parameter $M_{\mathrm{eff}}$ numerically by requiring that the first minimum of $G_{\mathcal M}(\tau;M_{\mathrm{eff}})$ equal the independently fitted value $C f_m^q$.
Writing $\tau_B$ for the location of this minimum, the predicted trajectory is
\begin{equation}
 \widehat{\xi^2}(\tau)
 =G_{\mathcal M}(\tau;M_{\mathrm{eff}}),
 \qquad 0\le\tau\le\tau_B,
 \label{eq:v:trajectory}
\end{equation}
with $\widehat{\xi^2}(\tau_B)=C f_m^q$ and $\widehat t_{\min}=\tau_B/\Gamma_0$; no timing correction is fitted.

\begin{figure}[tbp]
 \centering
 \includegraphics[width=\columnwidth]{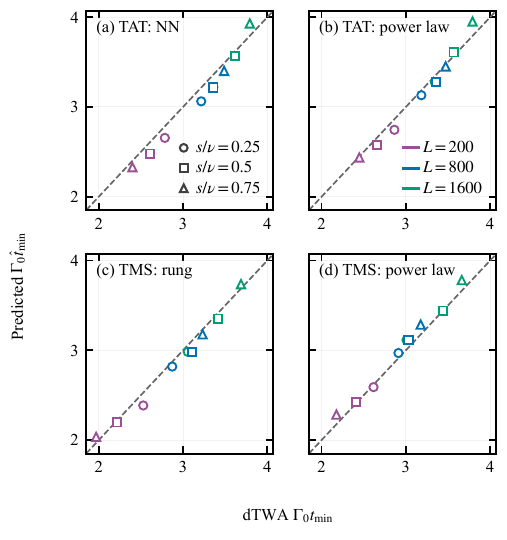}
\caption{Predicted and dTWA times of minimum squeezing at $\alpha=3$, expressed as $\Gamma_0t_{\min}$.
Colors distinguish $L=200,800,1600$, and marker shapes distinguish the three choices of $s/\nu$.
Predicted times are the first minima of the matched analytical trajectory, Eq.~\eqref{eq:v:trajectory}, with its minimum squeezing fixed by the independently fitted law $C f_m^q$.
Dashed lines indicate equality.}
 \label{fig:v:timeparity}
\end{figure}

Figures~\ref{fig:v:trajectories} and~\ref{fig:v:timeparity} compare these predictions with the dTWA dynamics and minimum times.
The preparation-time optimization of Sec.~\ref{sec:time} uses the same trajectory up to its first minimum.
We additionally allow the unevolved product state, i.e., we minimize $\min\{1,\widehat{\xi^2}_{\mathrm{budget}}\}$; this matters only at very short times, where $G_{\mathcal M}$ differs from unity by $O(M_{\mathrm{eff}}^{-2})$.
For each model, $\chi$ is optimized over $\mathcal I_L=[\chi_{\min},\chi_{\max}(L)]$, including strengths between the simulated values.
The lower bound is common to all sizes and equals the weakest protection on the $L=200$ paths with $s/\nu=1/4,1/2,3/4$.
For $L=200,800,1600$, $\chi_{\max}(L)$ is the largest protection strength on these three paths.
At $L=400$, where all paths pass through the common reference point, the upper bound is set by an additional simulation with $f_m=0.015$. 

\subsection{Collective reference at the stability threshold}
\label{app:v:collective_reference}

We evaluate the stability threshold $\chi_c(L)$ of Eq.~\eqref{eq:critical_protection} over all allowed nonzero momenta, using the pairing and protection spectra of each spatial model.
The first collective Wineland minimum $(\xi^2_{\min})^{\mathrm{coll}}(L)$ and its simulation time $t_{\min}^{\mathrm{coll}}(L)$ are obtained from the exact finite-$L$ symmetric-spin dynamics described in Appendices~\ref{app:v:tmscollective} and~\ref{app:v:tatcollective}.
Assigning the threshold protection cost gives the reference preparation time
\begin{equation}
 \mathcal T_{\mathrm{ref}}(L)
 =[3\chi_c(L)+2]\,t_{\min}^{\mathrm{coll}}(L).
 \label{eq:v:collective_reference}
\end{equation}
The stars in Figs.~\ref{fig:v:exactL} and~\ref{fig:v:alpha2} mark $(\mathcal T_{\mathrm{ref}},(\xi^2_{\min})^{\mathrm{coll}})$ and $(\mathcal T_{\mathrm{ref}},\chi_c)$ in the upper and lower rows, respectively.

At $\chi_c$, the critical finite-momentum modes are marginal; strict linear stability requires $\chi>\chi_c$.
Stable modes can nevertheless be excited, so suppressing their instabilities does not establish ideal collective dynamics.
The stars provide collective dynamical references at the threshold cost; attaining their squeezing in the spatial models may require stronger protection and longer preparation times.

\subsection{Optimal protection at fixed preparation time for \texorpdfstring{$\alpha=2$}{alpha=2}}
\label{app:v:alpha2}

Section~\ref{sec:time} considered dipolar Heisenberg interactions, $W_{ij}=r_{ij}^{-3}$.
Here we consider longer-range protection, $W_{ij}=r_{ij}^{-2}$, in the four one-dimensional models.
We use the same matched analytical trajectory and preparation-time optimization as in Sec.~\ref{sec:time}, with $C$ and $q$ fitted to the $\alpha=2$ data for each model.
Here the common lower bound is the weakest protection in the $L=800$ paths with $s/\nu=1/4,1/2,3/4$.
The upper bound at each size is the largest protection strength in these same three paths.

\begin{figure*}[tbp]
 \centering
 \includegraphics[width=\textwidth]{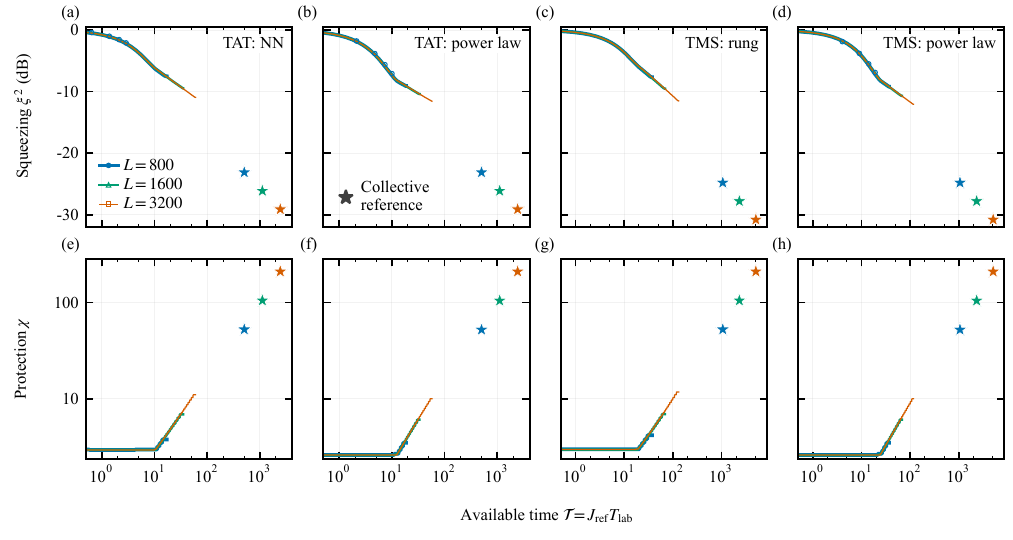}
\caption{Squeezing at fixed preparation time for $\alpha=2$ and $L=800,1600,3200$.
(a)--(d) Curves show the smallest predicted Wineland squeezing parameter $\xi^2$ attainable within the available time $\mathcal T=J_{\mathrm{ref}}T_{\mathrm{lab}}=(3\chi+2)t$.
Open markers show dTWA at the predicted optimal protection $\chi_\star$ and stopping time, wherever $\chi_\star$ coincides with a simulated protection strength.
(e)--(h) Corresponding optimal protection strengths.
Each curve ends when the lowest predicted squeezing within the sampled protection range is first reached.
Stars show ideal collective references at the stability-threshold cost defined in Appendix~\ref{app:v:collective_reference}.}
 \label{fig:v:alpha2}
\end{figure*}

Figure~\ref{fig:v:alpha2} shows that the preparation-time trade-off found for dipolar protection persists at $\alpha=2$: weak protection remains optimal at short times and stronger protection takes over at longer times, while the smaller dispersion exponent ($\nu=1$ instead of $\nu=2$) reduces both the protection threshold $\chi_c(L)$ and the collective reference times relative to $\alpha=3$.

\end{document}